\documentclass[conference]{IEEEtran}
\IEEEoverridecommandlockouts
\usepackage{cite}
\usepackage{authblk}
\usepackage{booktabs}
\usepackage{bm}
\usepackage{amsmath}
\usepackage{amssymb}
\usepackage{flushend}
\usepackage{setspace}
\usepackage{cases}
\usepackage{enumitem}
\usepackage{varwidth}
\usepackage{makecell}
\usepackage{url}
\usepackage{comment}
\usepackage{array}
\usepackage[ruled]{algorithm2e} 
\usepackage{graphicx}
\usepackage{balance}
\usepackage{multirow}
\usepackage{lipsum}
\usepackage{capt-of}
\usepackage{epstopdf}
\usepackage{comment} 
\usepackage[caption=false]{subfig}
\usepackage{dsfont}
\usepackage{xcolor}

\newlength\myindent
\newcommand{\n}{FIS-ONE}
\newcommand{\nn}{RF-GNN}

\allowdisplaybreaks

\newtheorem{theorem}{Theorem}

\begin{document}
\title{\n{}: Floor Identification System with One Label for Crowdsourced RF Signals\vspace{-0.1in}}
\author[$\ast$]{Weipeng Zhuo}
\author[$\ast$]{Ka Ho Chiu}
\author[$\ast$]{Jierun Chen}
\author[$\ast$]{Ziqi Zhao}
\author[$\ast, \S$]{S.-H. Gary Chan}
\author[$\dagger$]{Sangtae Ha}
\author[$\ddagger, \S$]{Chul-Ho Lee\thanks{\hspace{-0.1in}\noindent This work was supported in part by Hong Kong General Research Fund (under grant number 16200120). The work of Chul-Ho Lee was supported in part by the NSF under Grant IIS-2209921. \vspace{1mm}\\$^\S$Corresponding authors.}}

\affil[$\ast$]{The Hong Kong University of Science and Technology}
\affil[ ]{\textsuperscript{$\dagger$}University of Colorado Boulder, \textsuperscript{$\ddagger$}Texas State University}
\affil[ ]{Email: \textsuperscript{$\ast$}\{wzhuo,khchiuac,jcheneh,zzhaoas,gchan\}@ust.hk, \textsuperscript{$\dagger$}sangtae.ha@colorado.edu, \textsuperscript{$\ddagger$}chulho.lee@txstate.edu
\vspace{-0.25in}
}
\maketitle

\begin{abstract}
Floor labels of crowdsourced RF signals are crucial for many smart-city applications, such as multi-floor indoor localization, geofencing, and robot surveillance. To build a prediction model to identify the floor number of a new RF signal upon its measurement, conventional approaches using the crowdsourced RF signals assume that at least few labeled signal samples are available on each floor. In this work, we push the envelope further and demonstrate that it is technically feasible to enable such floor identification with only \emph{one floor-labeled} signal sample on the bottom floor while having the rest of signal samples \emph{unlabeled}.

We propose \n{}, a novel \textbf{f}loor \textbf{i}dentification \textbf{s}ystem with only \textbf{one} labeled sample. \n{} consists of two steps, namely signal clustering and cluster indexing. We first build a bipartite graph to model the RF signal samples and obtain a latent representation of each node (each signal sample) using our attention-based graph neural network model so that the RF signal samples can be clustered more accurately. Then, we tackle the problem of indexing the clusters with proper floor labels, by leveraging the observation that signals from an access point can be detected on different floors, i.e., signal spillover. Specifically, we formulate a cluster indexing problem as a combinatorial optimization problem and show that it is equivalent to solving a traveling salesman problem, whose (near-)optimal solution can be found efficiently. We have implemented \n{} and validated its effectiveness on the Microsoft dataset and in three large shopping malls. Our results show that \n{} outperforms other baseline algorithms significantly, with up to 23\% improvement in adjusted rand index and 25\% improvement in normalized mutual information using only one floor-labeled signal sample.

\end{abstract}

\section{Introduction} 
\label{sec:intro}
Many smart-city applications are enabled by radio frequency (RF) signals with floor labels. Such applications include multi-floor navigation in cities~\cite{luo2017mpiloc,tan2021implicit,gao2022federated,wang2023leto}, geofencing for pandemic control~\cite{helmy2016alzimio,megges2017technology,monir2021iot,zhuo2022semi}, robot rescue or navigation in environments where visual information is not available~\cite{imdoukh2017semi,walter2022extinguishing}, and unmanned aerial vehicle surveillance in restricted areas~\cite{hermand2018constrained,queralta2020uwb,vanhie2021indoor}. In these scenarios, it is costly and labor-intensive to employ trained surveyors to collect all the RF signals with floor labels. One practical solution is to leverage \emph{crowdsourcing}, where different people contribute different subsets of signals collected in a building. However, the crowdsourced RF signals, albeit abundant to cover the whole building, are largely unlabeled. Hence, it is important how to leverage the \emph{unlabeled} RF signals for floor identification.

\begin{figure}[t]
    \centering
    	\subfloat[]{%
        \includegraphics[width=0.33\linewidth]{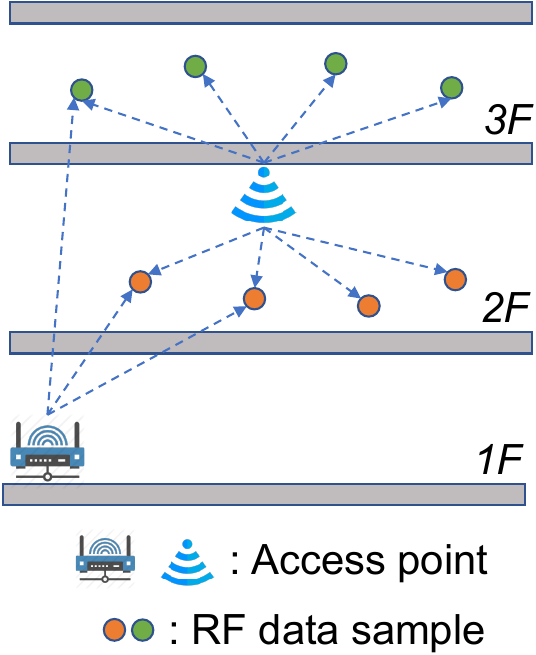}
        }
    \hspace{0.1in}
    \subfloat[]{%
        \includegraphics[width=0.55\linewidth]{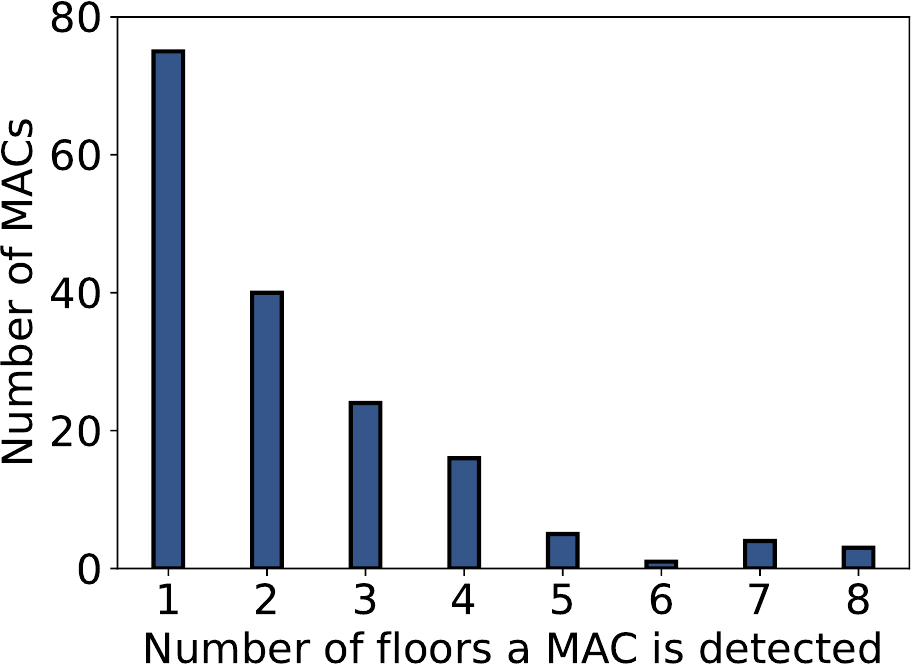}
        }
    	\caption{(a) An illustrative example of signal spillover. (b) Adjacent floors observe a more significant spillover effect.}
    	\label{fig:spill_over}
     \vspace{0.05in}
\end{figure} 

\begin{figure*}[t]
    \centering
    \includegraphics[width=0.9\textwidth]{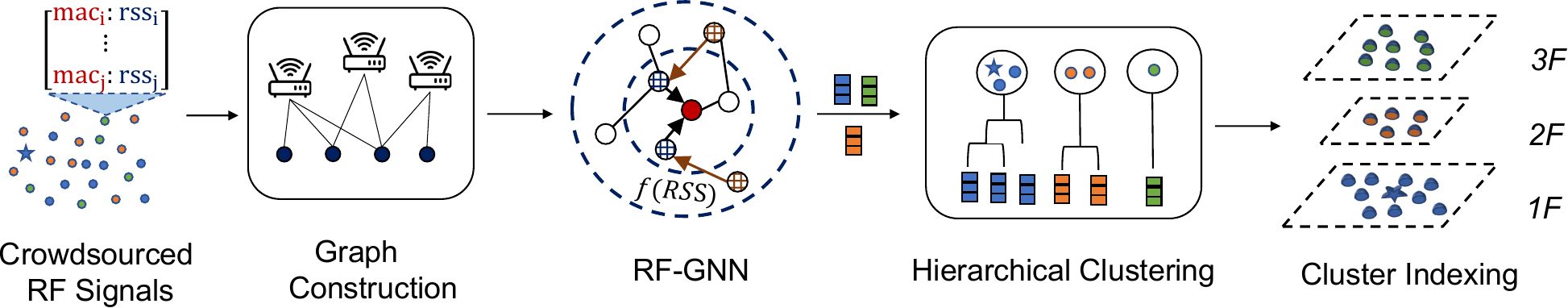}
    \caption{System overview of \n{}.}
    \label{fig:system_fis}
    \vspace{-0.2in}
\end{figure*}

Traditionally, different sensors, such as barometers and inertial measurement units (IMUs), have been leveraged to detect floor changes~\cite{luo2017mpiloc,bisio2018wifi,li2018multi,elbakly2018hyrise,ashraf2019floor,hao2020multi,yu2020precise,yang2020transloc,he2022tackling}, but these techniques either suffer from device heterogeneity or require users to follow specific routes for data collection. There have also been other studies which explore signal propagation models~\cite{elbakly2018truestory,elbakly2018hyrise,caso2019vifi,elbakly2020storyteller,shao2021floor} to predict floor labels. However, the locations of access points (APs) are required in the studies, hindering their solutions from being deployed in practice. Recently, there is a growing interest in developing machine learning-based solutions~\cite{dou2020bisection,wang2021secure,yan2021extreme,GRAFICS,chen2023run,gao2022federated,zhang2022towards} for floor identification due to their strong learning capability and high prediction accuracy. They, however, need to train models with a substantial amount of labeled data, which are difficult to obtain in crowdsourcing scenarios. Such a strong requirement of labeled data greatly hampers the large-scale deployment of the aforementioned applications using crowdsourced RF signals.

One natural question is \emph{how much we can eliminate the need of such expensive floor-labeled RF signals}. In this work, we demonstrate that it is technically feasible to infer floor labels of RF signals (upon their arrival) \emph{just using only one labeled signal sample on the bottom floor} while the rest of crowdsourced signal samples are \emph{unlabeled}. Specifically, we are able to eliminate the need of floor-labeled signal samples \emph{significantly} by leveraging the `signal spillover' effect. As shown in Figure~\ref{fig:spill_over}(a), a transmitted signal from an AP can be detected across different floors, i.e., the signal spills over to different floors. Intuitively, adjacent floors would see more and stronger signals from each other than distant floors, i.e., having a higher signal spillover effect between adjacent floors. This is validated in Figure~\ref{fig:spill_over}(b), where we show the number of common APs, or, more precisely, media access control (MAC) addresses that are detectable across different floors in a large shopping mall, i.e., a building of eight floors, where there are a total of 168 MAC addresses detected. For instance, if a MAC address can be detected across four floors, it will only be counted once in the bin of ``4'' in Figure~\ref{fig:spill_over}(b). We see that signals of most APs can spill over to neighboring floors. Note that a few MACs could be detected in many floors because there is a large empty space in the middle of the mall. 

Given this spillover observation, we expect that if we are able to group signals from the same floor together and figure out which groups are direct neighbors based on the signal spillover, then all the groups can be ordered, i.e., direct neighbor groups are placed next to each other. Since there is also a labeled signal sample on the bottom floor, the cluster containing the labeled sample is considered as the one for the starting floor, thereby making the ordering complete for floor identification. Thus, we propose \n{}, a novel \textbf{f}loor \textbf{i}dentification \textbf{s}ystem based on crowdsourced RF signals, where only \textbf{one} labeled signal sample is needed from the bottom floor. As illustrated in Figure~\ref{fig:system_fis}, \n{} consists of the following steps: The crowdsourced RF signals are first modeled as a bipartite graph, which is then processed by our radio-frequency graph neural network (RF-GNN) to obtain their vector representations (embeddings). These vector representations are further grouped into different clusters whose number is the same as the number of floors. Finally, the clusters are indexed properly.

To cluster the crowdsourced RF signals, we first model RF signals as a bipartite graph. RF signals are inherently heterogeneous, meaning that different signal samples would observe different subsets of APs in the building, even if they are collected on the same floor. Thus, it may not be feasible to use a vector of the superset of APs from the building to represent each signal sample, as there would be many missing entries in each vector, which could make clustering inaccurate. With the bipartite graph modeling, APs, or, more specifically, MAC addresses, are considered nodes of one type, i.e., MAC nodes, and signal samples are considered nodes of the other type, i.e., signal-sample nodes. A MAC node and a signal-sample node are connected if the MAC address is detected in the signal sample.

We then obtain a vector representation (or embedding) of each node with a graph neural network model. High-quality vector representations of the signal-sample nodes can preserve the relative distance (similarity) among the signal samples in the embedding vector space, i.e., if two signal samples are similar to each other in the physical space, their vector representations are also close to each other in the embedding space. Graph embedding techniques~\cite{LINE,gao2018bine,GRAFICS} can be used to obtain such representations, but they are limited to static bipartite graphs. In other words, they are not a suitable choice for dealing with new incoming RF signals, i.e., new nodes into the graph. To enable efficient representation learning on such a dynamic bipartite graph with incoming nodes (new RF signals), in this work, we design \nn{}, an attention-based graph neural network (GNN) model for RF signals. Specifically, \nn{} incorporates received signal strength (RSS) between a MAC node and each of its connected signal-sample nodes as a special type of attention, such that node representations can be learned effectively. With the learned representations of signal-sample nodes, we then apply the hierarchical clustering algorithm to divide them into a given number of floor clusters accurately.

After obtaining the clusters, we index the clusters, i.e., identifying which cluster corresponds to which floor, by leveraging the signal spillover effect. We first propose a novel measurement metric to measure the similarity between clusters based on the level of the signal spillover. The higher the spillover level, the closer the two clusters are. We next formulate the cluster indexing problem as a combinatorial optimization problem, which is to find an optimal ordering of clusters such that the spillover level between any two adjacent clusters is maximized. We show that it is equivalent to solving a travelling salesman problem (TSP), more specifically the problem of finding the shortest Hamiltonian path. Given the spillover levels between pairwise clusters (cities), it boils down to finding an optimal path that visits each cluster (city) exactly once such that the sum of the spillover levels (distances) is maximized (minimized). Since we have one labeled signal sample on the bottom floor, the cluster with the labeled data sample is treated as the starting cluster (city). We empirically validate that the visiting sequence in the optimal path accurately indexes the clusters with floor numbers.

We further discuss how \n{} can be extended to the case when the one labeled signal sample comes from an \emph{arbitrary} floor. This randomness would make the starting point of the TSP unfixed, leaving numerous paths (orderings) as candidate solutions. In other words, if the labeled signal sample comes from a different floor than the bottom one, the cluster containing the labeled sample can no longer be used as the starting cluster for the TSP, so the solutions to the TSP with different starting clusters need to be all evaluated. Thus, we propose a simple yet efficient heuristic method and numerically demonstrate that it still achieves accurate floor identification without much performance degradation ($\sim $3\%).

Our contributions can be summarized as follows.

\begin{itemize}
    
    \item \emph{\n{}: a novel floor identification system with only one labeled signal sample for crowdsourced RF signals.} \n{} is able to infer the floor number of each crowdsourced RF signal with only one labeled signal sample from the bottom floor, which greatly reduces the label requirement for the floor identification system and allows us to take a first step towards \emph{unsupervised} floor identification for crowdsourced RF signals.
    
    \item \emph{\nn{}: a novel attention-based graph neural network model to process heterogeneous RF signals.} \nn{} enables efficient representation learning on the graph built by RF signals by incorporating the RSS values as a type of attention to encode different levels of importance over edges, so that the vector representation of each signal-sample node is learned more accurately.
    
    \item \emph{Cluster indexing based on the signal spillover effect.} We index the clusters with proper floor numbers based on our observation of the signal spillover effect between floors. To achieve high indexing accuracy, we propose a novel measure of similarity between clusters depending on the level of the signal spillover effect and then solve a cluster indexing problem, which is transformed into a TSP, to obtain the optimal indexing, i.e., floor identification of unlabeled signal samples.
    
    \item \emph{Extensive experiments on two large-scale crowdsourced datasets.} We implement \n{} and evaluate its performance using Microsoft open dataset and in three large shopping malls. Experiment results show that \n{} achieves high accuracy for all buildings using only one floor-labeled signal sample on the bottom floor. \n{} outperforms other baseline algorithms significantly with up to 23\% improvement in adjusted rand index and 25\% improvement in normalized mutual information.
    
\end{itemize}

\section{Related Work}
\label{sec:related}
\noindent \textbf{Requirement of a substantial amount of labeled data:} A substantial number of floor-labeled RF signal samples are required in existing floor identification systems~\cite{dou2020bisection,wang2021secure,yan2021extreme,GRAFICS,gao2022federated,zhang2022towards} which are purely based on RF signals. For instance, in~\cite{yan2021extreme}, a floor-level classifier is first trained using labeled RF signals collected from different floors in a building before its online deployment. RMBMFL~\cite{wang2021secure} selects reliable APs and extracts features from RF signals coming from the APs to train a softmax classifier with corresponding floor labels. GRAFICS~\cite{GRAFICS} assumes that a few labeled RF signals are available on \emph{every} floor for floor identification. FedDSC-BFC~\cite{gao2022federated} obtains a collection of datasets of floor-labeled RF signals from different sensing clients in a crowdsourced manner and trains a floor classification model with federated learning. In contrast, \n{} aims to go beyond the conventional assumption on the presence of floor-labeled RF signals on every floor and demonstrates the feasibility of floor identification with only one labeled RF signal sample on the bottom floor while the rest of samples are unlabeled.

\vspace{1mm}
\noindent \textbf{Requirement of AP locations:} AP locations are necessary for other RF signal-based floor identification systems~\cite{elbakly2018truestory,elbakly2018hyrise,caso2019vifi,elbakly2020storyteller,shao2021floor} that do not require floor labels. For instance, HyRise~\cite{elbakly2018hyrise} measures, in an offline phase, the pressure readings of RF signals and then obtains the AP floor information using the pressure readings. These information are stored in a database for online inference. StoryTeller~\cite{elbakly2020storyteller} first identifies APs with the strongest signals among the measured RF signals and then converts the signal distribution into images with corresponding AP locations. These images are used to train a convolutional neural network model for floor classification. However, the locations of APs are generally difficult to obtain in practice, especially in the crowdsourcing scenarios. \n{} leverages only RF signal readings and does not require such AP locations during the floor identification process.

\vspace{1mm}
\noindent \textbf{Requirement of other sensors:} Other sensor signals~\cite{luo2017mpiloc,bisio2018wifi,li2018multi,elbakly2018hyrise,ashraf2019floor,hao2020multi,yu2020precise,yang2020transloc} have also been used to facilitate the floor identification process. In~\cite{bisio2018wifi}, it is observed that slow updates of pressure readings may be caused by reasons such as weather changes, while sudden changes of pressure readings are due to user movement. This observation is then leveraged by setting a threshold on the pressure readings to detect floor changes. However, it is usually difficult to set the threshold accurately in practice. 3D-WFBS~\cite{yu2020precise} learns relative altitude information from barometer readings and then obtains absolute floor information by combining the barometer reading and RSS from landmark APs. The deployment of landmark APs is, however, still challenging. MPiLoc~\cite{luo2017mpiloc} takes advantage of trajectories generated by IMU signals and then uses a barometer to separate the trajectories into different floors. However, the collection of IMU signals often incurs significant overhead in data storage for mobile devices. In contrast, \n{} is able to achieve high accuracy in floor identification based only on crowdsourced RF signals, among which a signal sample obtained on the bottom floor only needs to be floor-labeled.

\section{\nn{}: Attention-based Graph Neural Networks for RF Signals}
\label{sec:sage}
With crowdsourced RF signal samples, we first model them as a bipartite graph and then process the graph using our \nn{} to obtain a vector representation of each node.

\subsection{\nn{}: Graph Construction}
\label{subsec:bipartite}
\begin{figure}[t]
    \centering
    \includegraphics[width=0.48\textwidth]{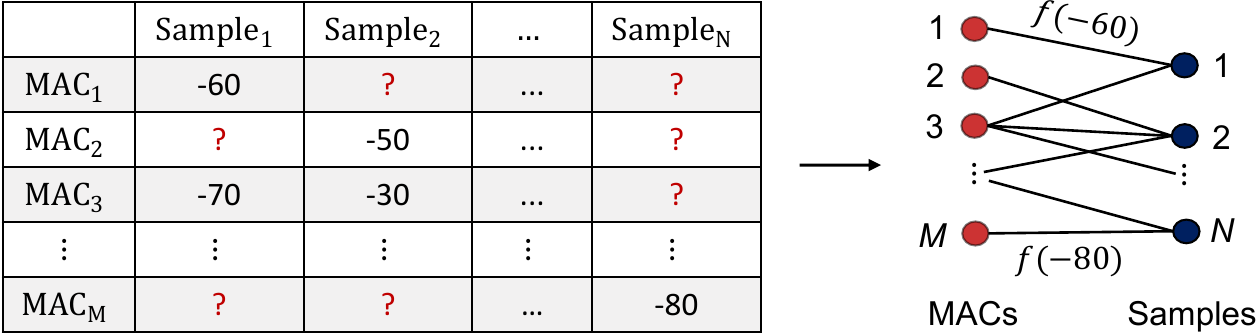}
    \caption{Graph modeling of RF signal samples does not have the problem of missing values by matrix modeling (unit: dBm).}
    \label{fig:missing_value}
    \vspace{0.1in}
\end{figure}

RF signals are heterogeneous, meaning that different RF signals may only observe different subsets of APs in the building. A traditional way to represent an RF signal is to use a vector consisting of the \emph{superset} of all APs in the building, which makes many entries empty, as illustrated in Figure~\ref{fig:missing_value}. Such missing entries, which are typically filled up with some arbitrary small values, would lead to unsatisfactory application performance. Recently, RF signals are modeled as a bipartite graph~\cite{GRAFICS} to overcome the missing value problem. We adopt the bipartite graph modeling in \n{} and below explain its details for the sake of completeness.

As shown in Figure~\ref{fig:missing_value}, there are two types of nodes. One is for APs, or, more precisely, their MAC addresses, while the other is for RF signal samples (records). Recall that each RF signal sample contains a list of sensed MAC addresses along with their received signal strength (RSS) values. Then, a node of a MAC address is connected to another node corresponding to an RF signal sample if the MAC address is detected in the RF sample (record). Thus, we can represent the crowdsourced RF signals as a bipartite graph. Specifically, we construct a weighted bipartite graph $\mathcal{G} \!=\! (\mathcal{U}, \mathcal{V}, \mathcal{E})$, where $\mathcal{V}$ is the set of nodes representing the crowdsourced RF signal samples, $\mathcal{U}$ is the set of nodes representing the sensed MAC addresses, and $\mathcal{E}$ is the set of edges. Each edge $e_{uv} \in \mathcal{E}$ denotes the edge between $u \in \mathcal{U}$ and $v \in \mathcal{V}$. Let $\text{RSS}_{uv}$ be the RSS value of an RF signal from $u$ that appears in $v$. The edge weight $w_{uv}$ is then defined as $w_{uv} := f(\text{RSS}_{uv})$, where $f(\text{RSS}_{uv}) \!>\! 0$ for all $\text{RSS}_{uv}$. We use $f(\text{RSS}_{uv}) := \text{RSS}_{uv} + c$ for our weighted bipartite graph $\mathcal{G}$, where $c$ is some constant such that $c \!>\! \max\{|\text{RSS}_{uv}|, \forall u, v\}$. In our case, $c$ is set to 120~dBm.

\subsection{\nn{}: Vector Representation Learning for Nodes}
\label{subsec:network_train}
Next, we efficiently learn a vector representation of each node from the constructed graph. The advantage of learning high-quality representations is that the relative distance (similarity) between two `signal-sample' nodes in the physical space can be well preserved in the embedding vector space. We first elaborate on the general aggregation process~\cite{graphsage} in graph neural networks for representation learning and then introduce our proposed \nn{}. 

\begin{figure}[t]
    \centering
    \includegraphics[width=0.48\textwidth]{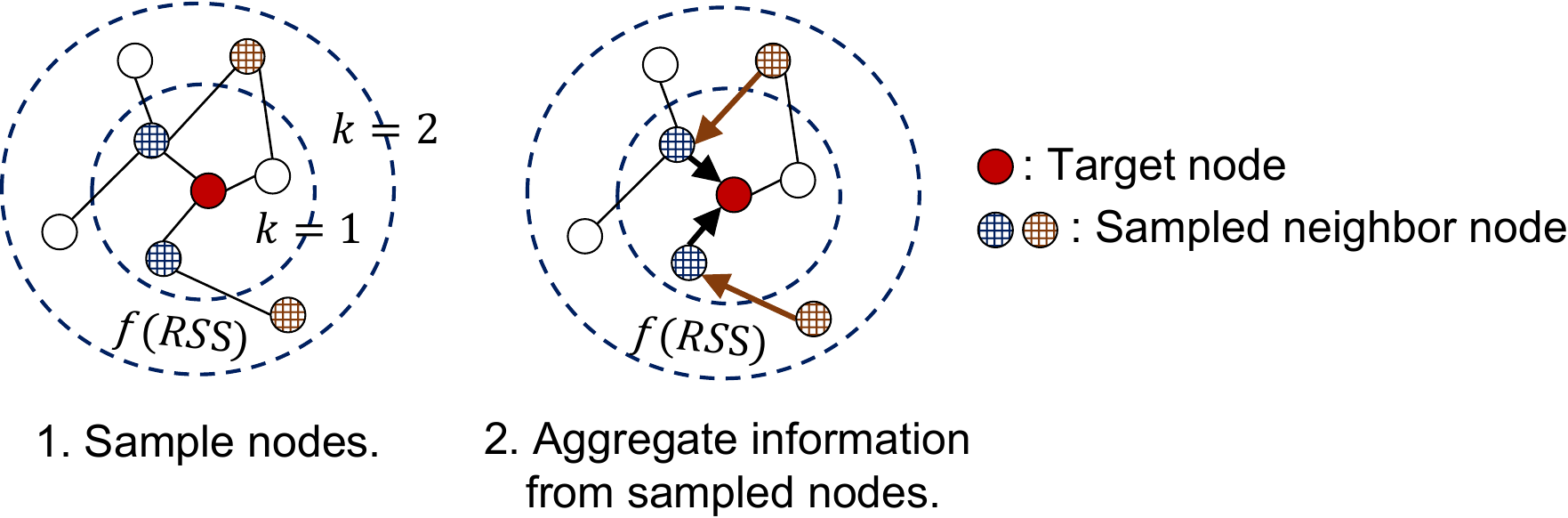}
    \vspace{-0.05in}
    \caption{An illustrative example of the general aggregation process. To learn the vector representation of a node, the graph neural network samples nodes from its first- and second-hop neighbors and then aggregates information from them. }
    \vspace{0.1in}
    \label{fig:aggregator}
\end{figure}

Given a target node whose representation is to be learned, there are two steps in the aggregation process. First, we sample nodes from the $N$-hop neighborhoods of the target node based on uniform distribution. Second, we aggregate information from the sampled nodes towards the target node. Figure~\ref{fig:aggregator} shows an illustrative example where the information is aggregated from two-hop neighbors towards the target node. It first samples two nodes from each of the first- and second-hop neighbors and implements two iterations of aggregation in the example. In each iteration, each node obtains information from its sampled immediate neighbors. After two iterations, the target node contains information from its two-hop neighbors. 

We below introduce our \nn{}, an attention-based GNN model for crowdsourced RF signals. In our scenario, it is natural to define the weights of edges as a function of sensed RSS values to encode different levels of signal strength from different APs (MAC addresses). To sample neighbors of a target node, intuitively, the higher the sensed RSS value between the node and its neighbor, the more likely the neighbor should be chosen. Thus, we design our own neighbor sampling strategy as follows. Consider $u \in U$ and $v \in V$ with $e_{uv} \in \mathcal{E}$ and suppose that $v$ is the target node. The sampling probability that $u$ is selected for aggregation is given by
\begin{equation}
Pr(u) = \frac{f(RSS_{uv})}{\sum_{u'\in N(v)} f(RSS_{u'v})}. \nonumber
\end{equation}
Similarly for when $u$ is the target node.

Graph attention networks have been introduced in the literature~\cite{velickovic2018graph,zhong2022tree} to learn \emph{better} representations by incorporating an attention mechanism into the aggregation process compared to the GNN model without attention, since they consider different neighbors with different levels of importance (attention weights). However, the weight learning process requires a substantial amount of labeled data for supervised or semi-supervised training, which is infeasible in our scenario. We instead observe that in our weighted bipartite graph, the edge weights naturally capture the importance of neighbors, i.e., higher edge weights (RSS values) generally indicate closer distances. Thus, we incorporate the edge weights as a type of attention into the aggregation process and design an aggregator based on edge weights. Specifically, let $N'(v)$ be the set of sampled neighbors of $v$ and let $\bm{r}_u$ be the vector representation of $u \in N'(v)$. The weighted aggregator is defined as
\begin{equation}
\mbox{AGGREGATE}^w=\sum_{u\in N'(v)}\frac{f(RSS_{uv})}{\sum_{u'\in N'(v)} f(RSS_{u'v})}\bm{r}_u. \nonumber
\end{equation}

We next explain the remaining details of \nn{} and its \emph{unsupervised} training to obtain the vector representation of each node. Consider $i \in U\cup V$. Let $\bm{r}_i^k$ be the representation of $i$ in the $k$-th iteration, let $N'(i)$ be the sampled neighborhood of $i$, and let $K$ be the number of hops. We set $\bm{r}_i^0$ to a random vector. In the $k$-th iteration of the aggregation process, \nn{} first aggregates information from its sampled direct neighbors and stores in a temporary variable, say, $\bm{r}_{N'(i)}^k$, which is given by
\begin{equation}
\bm{r}_{N'(i)}^k = \mbox{AGGREGATE}^w({\bm{r}_j^{k-1}, \forall j \in N'(i)}), \label{eqn:aggregate} \nonumber
\end{equation}
where $\bm{r}_{j}^{k-1}$ denotes the representation of neighbor $j$ in the $(k\!-\!1)$-th iteration. \nn{} then concatenates $\bm{r}_{N'(i)}^k$ with the vector representation of $i$ itself in the previous iteration, i.e., $\bm{r}_i^{k-1}$. The concatenated vector goes through a fully connected layer with a trainable weight matrix $\mathbf{W}^k$ and a non-linear function $\sigma(\cdot)$ to generate $\bm{r}_i^k$, which is given by
\begin{equation}
    \bm{r}_i^k = \sigma\left(\mathbf{W}^k \mbox{CONCAT}\left(\bm{r}_i^{k-1},\bm{r}_{N'(i)}^k\right)\right). \label{eqn:concat} \nonumber
\end{equation}
Finally, $\bm{r}_i^k$ is normalized as $\bm{r}_i^k := \bm{r}_i^k/||\bm{r}_i^k||_2$, where $||\cdot||_2$ is the $\ell_2$ norm. $\bm{r}_i^k$ is then used for the $(k\!+\!1)$-th iteration. After repeating the whole process $K$ times, the final representation for node $i$ is given by $\bm{r}_i^K$. 

For the process of unsupervised training, we follow the process in~\cite{graphsage}, which is commonly used in training GNN models in an unsupervised manner~\cite{you2020handling, zhang2019heterogeneous, yang2020multisage}. Specifically, it is based on a large number of short random walks whose length is of five steps generated on the graph. The intuition here is that the nodes that appear in the same random walk should have similar vector representations as they are close to each other. Suppose that nodes $i$ and $j$ co-occur in a short random walk and let $\bm{r}_i$ and $\bm{r}_j$ be their corresponding vector representations. We use the following loss function to learn the vector representation of each node and the weight matrices $\mathbf{W}^k$'s: 
\begin{equation}
    \mathcal{L}_{\mathcal{G}} := - \log \left( \sigma(\bm{r}_i\cdot\bm{r}_j)\right) - \tau\times \bm{E}_{z\sim \Pr(z)} \log \left( \sigma (-\bm{r}_i\cdot \bm{r}_z) \right), \nonumber
\end{equation}
where $\sigma(x) = 1/(1+\exp(-x))$, $\bm{r}_i \cdot \bm{r}_j$ denotes the inner product of $\bm{r}_i$ and $\bm{r}_j$, and the expectation $\bm{E}_{z\sim \Pr(z)}$ is with respect to $\Pr(z)$, which is a user-defined distribution over nodes. Note that the second term is based on the so-called `negative sampling' in that $\tau$ nodes are randomly sampled from the entire graph according to $\Pr(z)$, so they are less likely to appear in the same random walk. In other words, the first term encourages the nodes that co-occur in the same random walk to stay close to each other in the embedding vector space, while the second term forces the nodes that are probably far from each other to separate apart in the embedding vector space. As used in~\cite{Mikolov2013,LINE,ma2021deep}, we choose $\tau=4$ and $\Pr(z) \propto d_z^{3/4}$, where $d_z$ is the degree of node $z$.

\section{Signal Clustering and Cluster Indexing}
\label{sec:cluster_index}
After obtaining the latent vector representation of each node in the bipartite graph, we cluster the representations of signal-sample nodes into floor clusters and then index the clusters with proper floor numbers.

\subsection{Signal Clustering}
\label{subsec:clustering}

To cluster the representations of the signal-sample nodes into different clusters whose number is the same as the number of floors in the building, we employ a proximity-based hierarchical clustering. To start with, each representation is treated as a cluster. We merge two clusters with the shortest distance together in each round. Let $\bm{C}_i$ be the set of representations of the signal-sample nodes in cluster $i$. The distance between clusters $i$ and $j$ is then defined as
\begin{equation}
d(\bm{C}_i, \bm{C}_j) := \frac{1}{|\bm{C}_i| |\bm{C}_j|}\sum_{\bm{r} \in \bm{C}_i}\sum_{\bm{r}' \in \bm{C}_j} \| \bm{r} -\bm{r}' \|_2,\label{eqn:dist_cluster} \nonumber
\end{equation}
where $\bm{r} \in \bm{C}_i$ is the representation of a signal-sample node in cluster $i$. This clustering process continues until the number of clusters becomes the same as the number of floors in the building.

\subsection{Cluster Indexing: A TSP Formulation}
\label{subsec:indexing}

\begin{figure}[t]
    \vspace{1mm}
    \centering
    \includegraphics[width=0.48\textwidth]{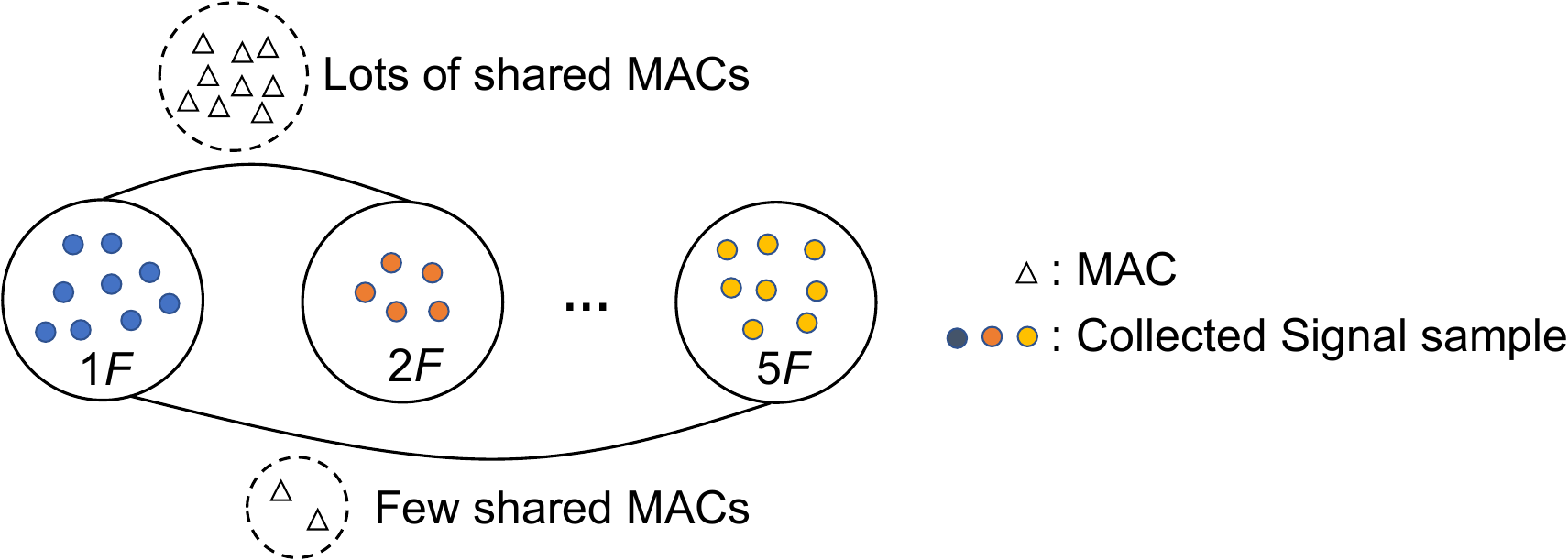}
    \caption{Adjacent floors are highly likely to detect more shared MACs due to the signal spillover.}
    \label{fig:jaccard_intuition}
    \vspace{0.2in}
\end{figure}

We next index the clusters with floor numbers. Recall that the signal from an AP can be detected on different floors, i.e., there is a signal spillover effect. Two adjacent floors would observe a higher spillover effect, as it was empirically validated in Figure~\ref{fig:spill_over}(b). See Figure~\ref{fig:jaccard_intuition} for an illustration. If we are able to infer which two clusters are direct neighbors based on the signal spillover effect, we can eventually obtain an ordering of the clusters. Here, since we have one floor-labeled signal sample measured from the bottom floor, we use the cluster including the labeled sample as the starting cluster in the ordering.

To that end, we need a measure of gauging the level of signal spillover effect between floors, which is now the similarity between their corresponding clusters. A natural choice here would be the Jaccard similarity coefficient~\cite{jaccard1912distribution} as a measure of similarity between two clusters, which becomes the ratio of the number of shared MACs to the total number of MACs detected in both clusters in our setting. To be precise, letting $\bm{A}_i$ be the set of MACs detected in cluster $i$, the Jaccard similarity coefficient $\bm{J}_{ij}$ for clusters $i$ and $j$ is given by
\begin{equation}
	\bm{J}_{ij} = \frac{|\bm{A}_i \cap \bm{A}_j|}{|\bm{A}_i \cup \bm{A}_j|}. \nonumber
\end{equation}
However, this measure only considers the presence of a MAC (a set element) rather than its coverage. For instance, there is no difference between a MAC that is sensed by most signal samples and another MAC that is only sensed by few signal samples in each cluster. The former would correspond to an AP that has a wider coverage than the latter, but such a difference cannot be captured by the Jaccard similarity coefficient.

To overcome this limitation, we propose an adapted Jaccard similarity coefficient to capture the coverage of each AP. Instead of simply measuring the existence of a MAC, we also consider its appearance frequency. Since crowdsourced RF signals are generally abundant, the frequency of a MAC that appears in a cluster (a collection of RF signals) should be a good indicator of its coverage. Consider two clusters $i$ and $j$, and suppose that there are a total of $m$ MACs detected in the clusters. Letting $f_{ik}$ be the frequency of MAC $k$ that appears in cluster $i$, we define the frequency count of \emph{shared} MACs between clusters $i$ and $j$ as
\begin{equation}
f^{\text{share}}_{ij} := \sum_{k=1}^{m} f_{ik}f_{jk}. \label{eqn:share}
\end{equation}
Here we do not simply compute the frequency of each MAC that appears in \emph{both} clusters $i$ and $j$ in which case we cannot see how its appearances are distributed over $i$ and $j$. For example, a MAC could appear predominately in one cluster over the other. Thus, we use the product of separate frequencies of each MAC in $i$ and $j$ for the frequency count $f^{\text{share}}_{ij}$. In addition, we define the frequency count of \emph{unshared} MACs between clusters $i$ and $j$ as
\begin{equation}
f^{\text{diff}}_{ij} := \sum_{k=1}^{m}\Big(\mathds{1}_{\{f_{ik}=0\}}  f_{jk} \bar{f}_i + \mathds{1}_{\{f_{jk}=0\}} f_{ik} \bar{f}_j \Big), \label{eqn:diff}
\end{equation}
where $\bar{f}_i$ is the \emph{average} frequency count of MACs appearing in cluster $i$, i.e., $\bar{f}_i \!=\! \sum_{k=1}^{m} f_{ik}/m$, and $\mathds{1}_{\{\cdot\}}$ is an indicator function. For example, $\mathds{1}_{\{f_{ik}=0\}}$ is given by
\begin{equation}
\mathds{1}_{\{f_{ik}=0\}} = 
\begin{cases}
	1, &\text{if MAC $k$ does not appear in cluster $i$,} \\
	0, &\text{otherwise}. \nonumber
\end{cases}
\end{equation}
Note that for the definition of $f^{\text{diff}}_{ij}$ in \eqref{eqn:diff}, we do not simply compute the \emph{pure} frequency count of unshared MACs between $i$ and $j$ as its value would not be on the same scale as that of $f^{\text{share}}_{ij}$ in \eqref{eqn:share}, which is in the product form. Thus, we consider $\bar{f}_i$ in the first term and $\bar{f}_j$ in the second term. From \eqref{eqn:share} and \eqref{eqn:diff}, our adapted Jaccard similarity coefficient $J^{n}_{ij}$ between clusters $i$ and $j$ is finally defined as
\begin{equation}
J^n_{ij} := \frac{f^{\text{share}}_{ij}}{f^{\text{share}}_{ij} + f^{\text{diff}}_{ij}}. \label{eqn:adapted}
\end{equation}
It is worth noting that this adapted Jaccard similarity coefficient makes the performance of floor identification better than the case with the original Jaccard similarity coefficient, as shall be demonstrated numerically in Section~\ref{sec:exp}.

For each pair of clusters, we calculate their adapted Jaccard similarity coefficients to gauge their similarity. The higher the coefficient is, the higher the similarity is, i.e., the clusters that correspond to adjacent floors should have higher coefficients than the ones corresponding to distant floors. Using the cluster that contains the only labeled signal sample as the starting cluster, \textbf{our cluster indexing problem is to find an optimal ordering of the clusters such that the sum of the pairwise (adapted Jaccard) coefficients of the clusters that are adjacent in the ordering is maximized.} Then, the optimal ordering simply indicates the floor number of each cluster, determining the labels of all the signal samples in the cluster.

\begin{figure}[t]
    \centering
    \includegraphics[width=0.98\linewidth]{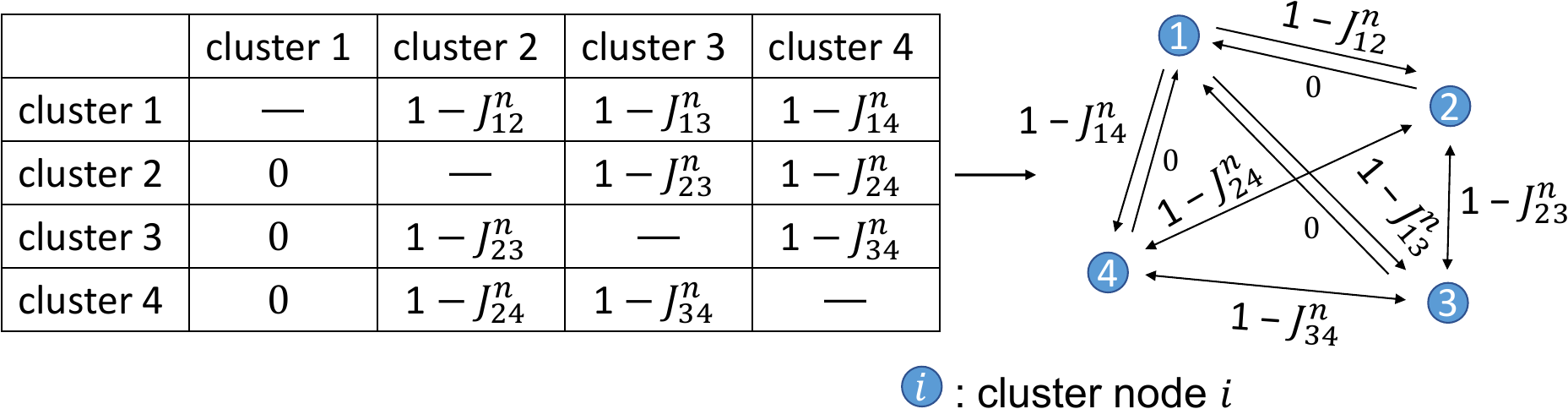}
    \caption{A complete graph formed by different clusters with their pairwise distances.}
    \label{fig:jaccard_to_graph}
    \vspace{0.05in}
\end{figure}

Consider a weighted complete graph $G$ with a set of nodes $\mathcal{N} \!=\! \{1, 2,\ldots, N\}$, where $N$ is the number of floors in a building of interest. Without loss of generality, suppose that node 1 corresponds to the cluster that contains the only labeled signal sample. Let $w_{ij}$ be an edge weight from nodes $i$ to $j$. Note that there is no self loop in $G$. We set the edge weight $w_{ij} := 1 - J^n_{ij}$ for all $i \in \mathcal{N}$ and $j \in \mathcal{N} \setminus \{1\}$, while setting $w_{i1} := 0$ for all $i \ne 1$. See Figure~\ref{fig:jaccard_to_graph} for an illustration. Note that $w_{ij} = w_{ji}$ for all $i,j \in \mathcal{N} \setminus \{1\}$ due to the symmetricity of $J^n_{ij}$ between $i$ and $j$. Then, we have the following.
\begin{theorem}
The cluster indexing problem is equivalent to solving a TSP variant, or finding the shortest Hamiltonian path, on $G$, which is formally given by 
\begin{align*}
\text{minimize~} & \sum_{i=1}^N\sum_{j=1}^N w_{ij}\mathds{1}_{ij} \\
\text{subject to~} & \sum_{i=1,i \neq j}^N \mathds{1}_{ij} = 1, \sum_{j=1,j \neq i}^N \mathds{1}_{ij} = 1, \text{~and~} \\
& \sum_{i,j \in \mathcal{S}, i \neq j} \mathds{1}_{ij} \leq |\mathcal{S}| - 1, ~\forall \mathcal{S} \subsetneq \mathcal{N}, |\mathcal{S}| \geq 2,
\end{align*}
where $\mathds{1}_{ij}$ is the indicator function, i.e.,
\begin{equation}
\mathds{1}_{ij} = 
\begin{cases}
	1, &\text{if the path goes directly from $i$ to $j$,} \\
	0, &\text{otherwise}. \nonumber
\end{cases}
\end{equation}
\end{theorem}
\begin{IEEEproof}
Recall that given a set of cities and the pairwise distances between cities, the TSP is to find the \emph{shortest} route that visits each city exactly once and returns to the starting city. If all the distances to the starting city are set to zero, it boils down to the problem of finding the shortest Hamiltonian path with a given starting city since the way back from the final city in the route does not contribute to the total route length. Also, observe from \eqref{eqn:adapted} that $0 \!\leq\! J^n_{ij} \!\leq\! 1$. Thus, with the settings of $w_{ij}$, the cluster indexing problem, which is a maximization problem, is equivalent to the problem of finding the shortest Hamiltonian path on $G$ starting with node 1. 
\end{IEEEproof}
\vspace{1mm}

Note that the exact solution to the TSP can be obtained by the Held-Karp algorithm~\cite{held1962dynamic} with the time complexity of $O(N^22^N)$. We can thus resort to the Held-Karp algorithm to solve our problem. Once the solution, i.e., the optimal ordering of the clusters, is obtained, the clusters are indexed with the corresponding floor numbers sequentially, with the first cluster being the bottom floor. In case $N$ is large, we can also resort to approximation algorithms~\cite{johnson1997traveling} to obtain the near-optimal ordering. We empirically validate in the next section that an approximation algorithm works reasonably well compared to the exact algorithm.

\section{Experiment Results} 
\label{sec:exp}
We present here the extensive experiment results for \n{}. We first discuss our experiment settings and compare the performance between \n{} and other baseline algorithms. We then study the impact of different system components and parameters on \n{}. Our code is available online.\footnote{https://github.com/SteveZhuo/FIS-ONE}

\begin{figure}[h]
    \centering
    \includegraphics[width=0.28\textwidth]{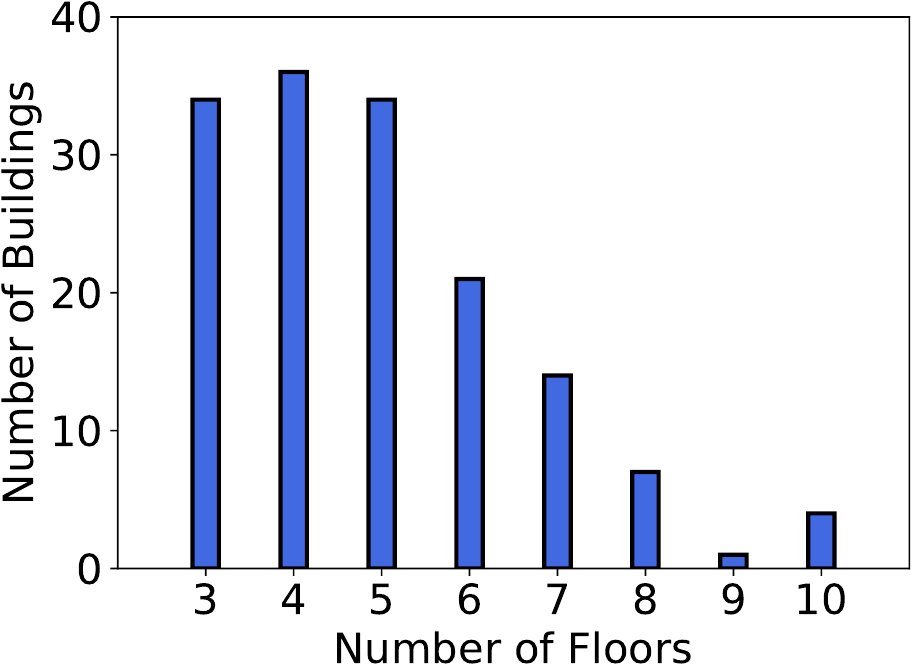}
    \vspace{-2mm}
    \caption{Distribution of the number of buildings (Two datasets combined).}
    \label{fig:building_num}
    \vspace{-0.1in}
\end{figure}

\subsection{Experiment Settings}
\label{subsec:exp_settings}

\noindent\textbf{Experiment setup:} We conduct experiments on the Microsoft's open dataset~\cite{ms-kaggle} (denoted as `Microsoft' in the results) and in three large shopping malls (denoted as `Ours' in the results). For the Microsoft dataset, we first filter out two-story buildings as we have one labeled signal sample on the starting floor, which makes the indexing straightforward. Since crowdsourced data are usually abundant, we also filter out floors with less than 100 RF signal samples while the other floors remain intact. We end up using the dataset of 152 buildings in which each \emph{floor} is associated with around 1000 RF signal samples on average, and the number of floors in a building ranges from three to ten. For the shopping malls, two of them have five floors while the other one has seven floors. We collected around 1000 RF signal samples on each floor. We show the floor number distribution of buildings in Figure~\ref{fig:building_num}. Unless otherwise mentioned, we present the average results of the buildings from each dataset. 

\vspace{1mm}
\noindent\textbf{Baseline algorithms for comparison:} To the best of our knowledge, there is \emph{no existing work} on floor identification with only one floor-labeled signal sample and the rest of the samples being unlabeled. Hence, we consider the following recent and popular clustering algorithms. Since they only provide clustering results (i.e., no cluster indexing) of RF signal samples, we adapt them with different components from \n{} such that they can be applied to our target scenario. Specifically, once we have the clusters generated by the baselines algorithms, we use our cluster indexing method explained in Section~\ref{sec:cluster_index} to label the resulting clusters with floor numbers. In addition, for SDCN~\cite{bo2020structural}, DAEGC~\cite{wang2019attributed} and METIS~\cite{karypis1998fast}, the bipartite graph constructed from RF signal samples is used as an input for them. On the other hand, for MDS~\cite{cox2008multidimensional}, a matrix representation of RF signal samples is used as an input, as illustrated in Figure~\ref{fig:missing_value}. The missing entries are filled with $-120$~dBm. To summarize, we have

\begin{itemize}[itemsep=2pt,leftmargin=1.5em]

    \item SDCN~\cite{bo2020structural}: It learns a vector representation of each node in the graph while at the same time grouping the representations into different clusters using a combination of a deep neural network model and a graph convolution network model.
    
    \item DAEGC~\cite{wang2019attributed}: It generates the embedding of each node in the graph using an autoencoder and gradually clusters the embeddings based on the cluster centroids that are being updated during training.

    \item METIS~\cite{karypis1998fast}: It is a popular graph partition algorithm, which first coarsens the graph and partitions the coarsened graph to obtain initial clusters. It then uncoarsens the graph to refine the clusters.

    \item Multidimensional scaling (MDS)~\cite{cox2008multidimensional}: It learns the embeddings of RF signal samples by using pairwise distances among the vectors of RF signal samples, which are represented in a matrix form. We here use the pairwise distance of 1 $-$ cosine similarity. The hierarchical clustering is then applied to the learned embeddings to obtain clusters.
\end{itemize}

\begin{table*}[t]
\caption{Performance comparison with baseline algorithms. The entry is in the form of mean (std).}
\centering 
\resizebox{0.8\textwidth}{!}{
\begin{tabular}{| c | c | c | c | c | c | c|}
\hline
\multirow{2}{*}{Algorithm}& \multicolumn{2}{c|}{$ARI$} & \multicolumn{2}{c|}{$NMI$}  & \multicolumn{2}{c|}{Edit Distance} \\
\cline{2-7}
& Microsoft & Ours & Microsoft & Ours & Microsoft & Ours \\
\hline
\n{} &   \textbf{0.856} (0.086) & \textbf{0.845} (0.012)   &  \textbf{0.878} (0.090) & \textbf{0.875} (0.017)  &  \textbf{0.880} (0.075) & \textbf{0.877} (0.006)  \\
\hline
SDCN & 0.714 (0.065) & 0.718 (0.023) & 0.751 (0.060) & 0.750 (0.036) & 0.820 (0.068) & 0.813 (0.012) \\
\hline
DAEGC & 0.697 (0.072) & 0.661 (0.032) & 0.776 (0.059) & 0.696 (0.016) & 0.808 (0.087) & 0.803 (0.021)\\
\hline
METIS & 0.641 (0.148) & 0.607 (0.028) & 0.657 (0.109) & 0.655 (0.012) & 0.760 (0.114) & 0.790 (0.017)\\
\hline
MDS & 0.622 (0.185) & 0.582 (0.049) & 0.708 (0.146) & 0.675 (0.048) & 0.790 (0.137) & 0.796 (0.015)\\
\hline
\end{tabular}
}
\label{tab:results_all}
\vspace{-0.0in}
\end{table*}

\begin{figure*}[t]
    \centering
     \includegraphics[width=0.55\linewidth]{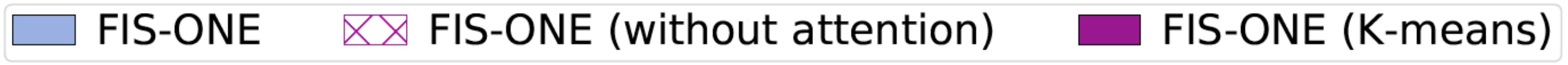} 
     
    \subfloat[Microsoft]{%
        \includegraphics[width=0.23\linewidth]{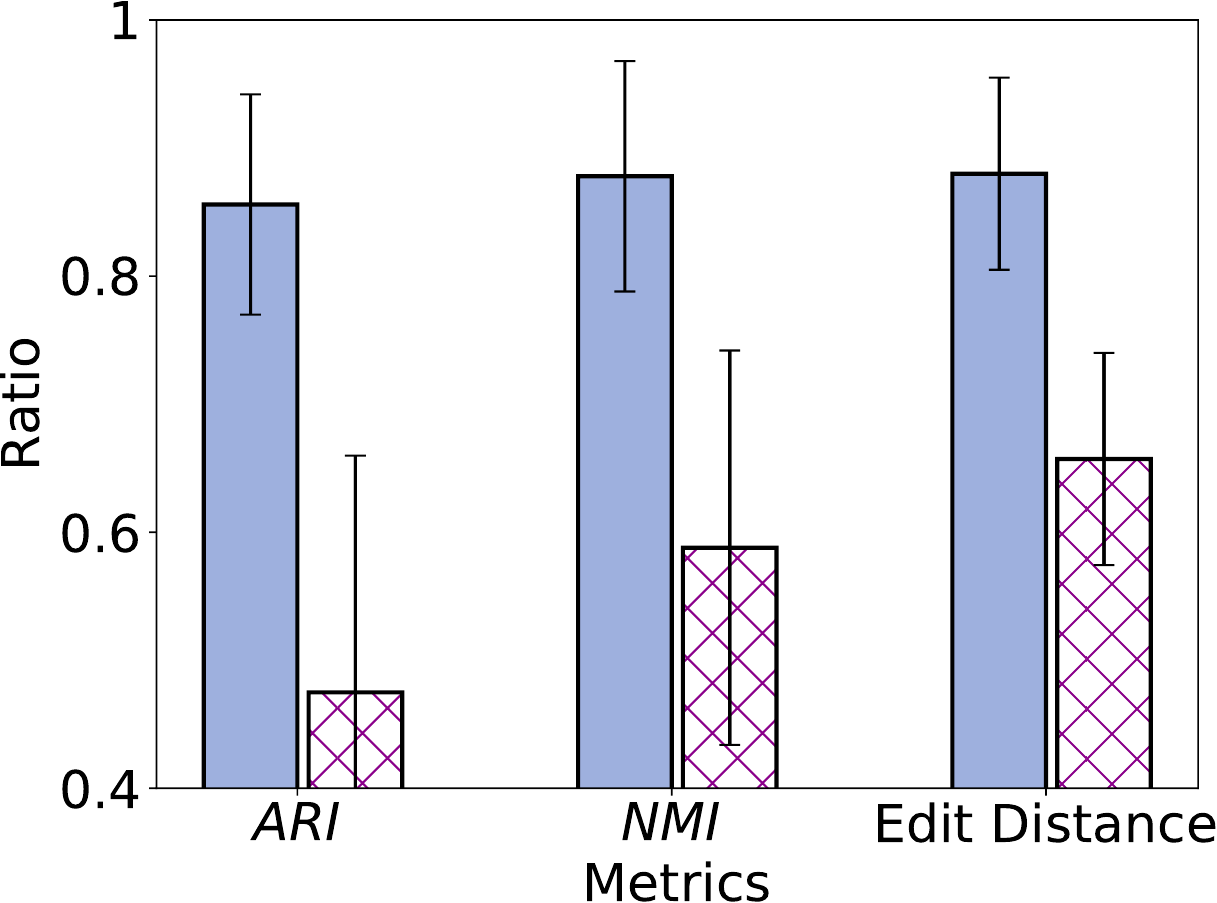}
        }    
    \hspace{0.05in}
    \subfloat[Ours]{%
        \includegraphics[width=0.23\linewidth]{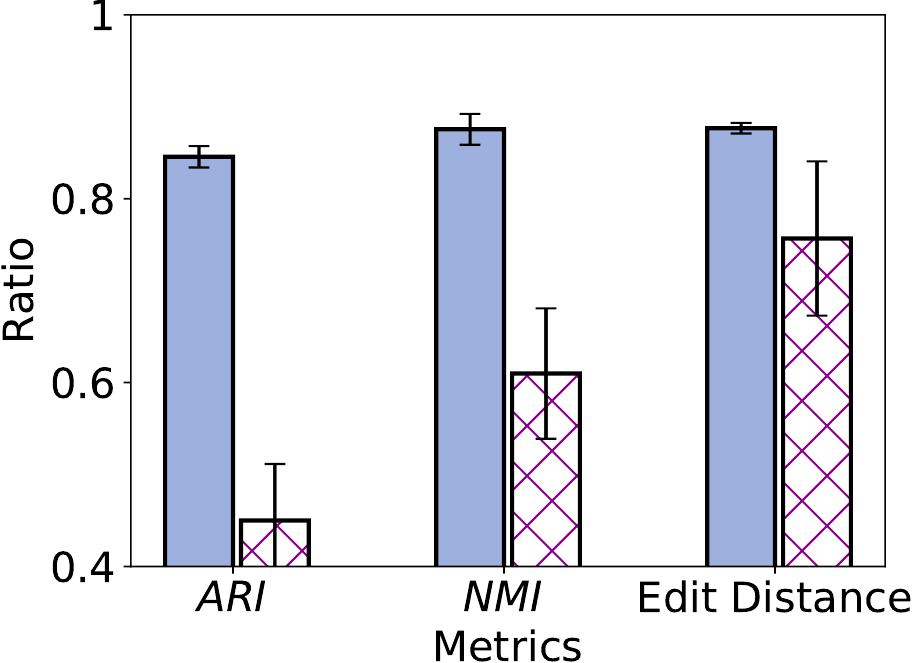}
        }    
    \hspace{0.05in}
    \subfloat[Microsoft]{%
        \includegraphics[width=0.23\linewidth]{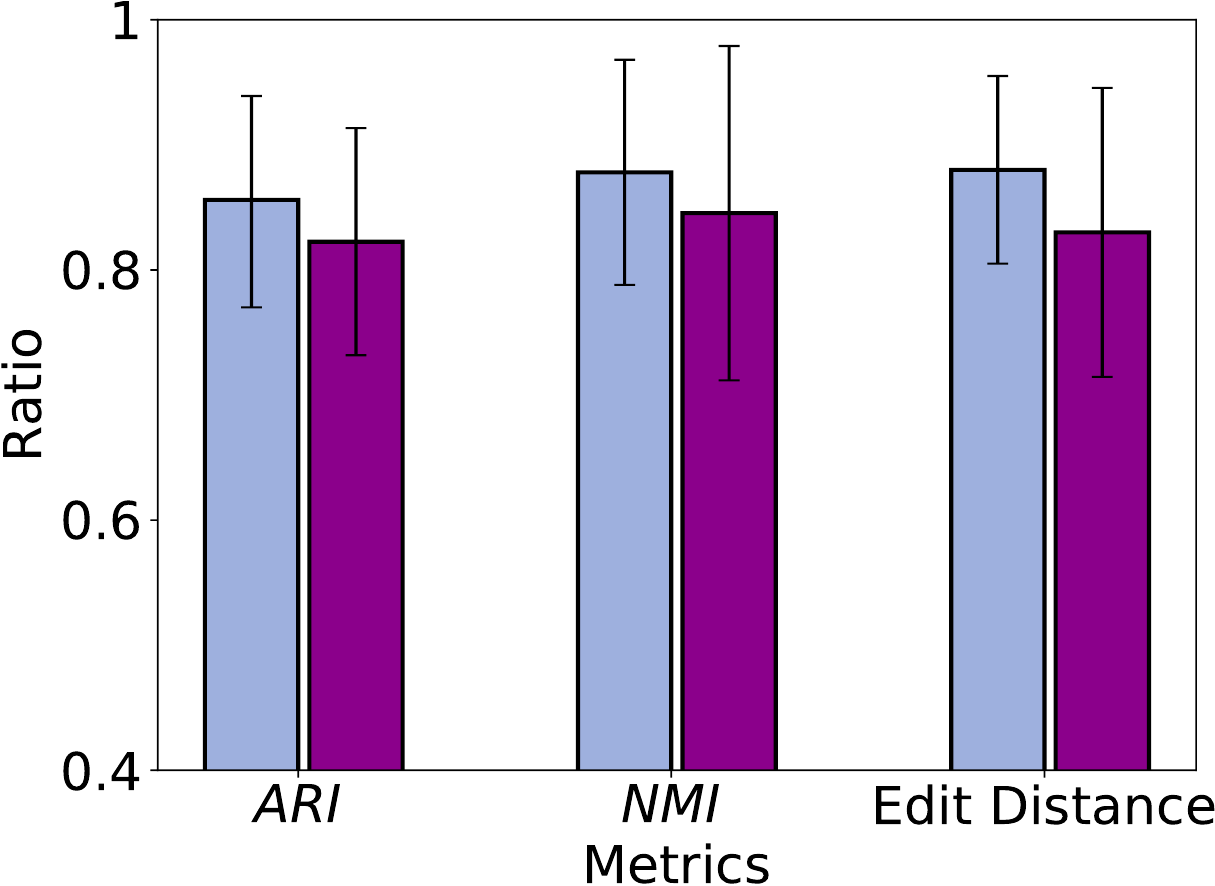}
        }
    \hspace{0.05in}    
    \subfloat[Ours]{%
        \includegraphics[width=0.23\linewidth]{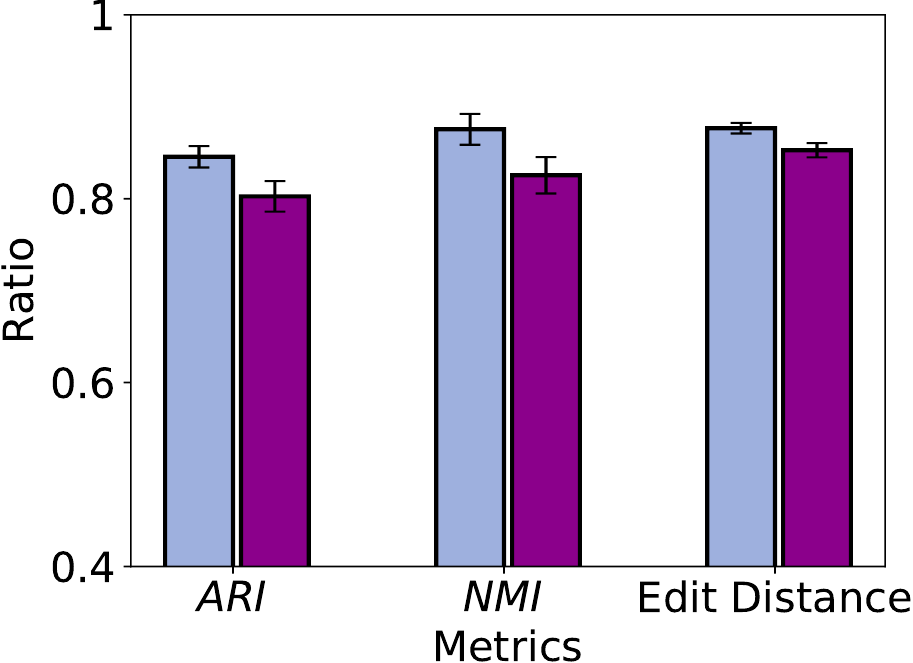}
        }
    	\caption{Ablation study of \n{}. (a) and (b) \n{} (without attention); (c) and (d) \n{} (K-means).}
    	\vspace{-0.1in}
    	\label{fig:ablation_ab}
\end{figure*} 

\noindent Note that the number of clusters obtained by each algorithm is the same as the number of floors in each building. For SDCN and DAEGC, we use their code provided in~\cite{bo2020structural} and~\cite{wang2019attributed}, respectively. We use the python implementation of METIS. 

\vspace{1mm}
\noindent\textbf{Evaluation metrics:} We use the adjusted rand index ($ARI$) and the normalized mutual information ($NMI$) to evaluate the clustering performance. Intuitively, $ARI$~\cite{rand1971objective} measures the pairwise data-point similarity between predicted clusters and ground-truth clusters. For instance, if two data points appear in the same cluster by both predicted clustering and ground-truth clustering, $ARI$ will be higher. Formally speaking, let $\bm{X} \!=\! (\bm{X}_1, \ldots, \bm{X}_N)$ be the predicted clustering results with corresponding clusters and let $\bm{Y} \!=\! (\bm{Y}_1, \ldots \bm{Y}_N)$ be the ground-truth clusters. Let $n_{ij} := |\bm{X}_i\cap \bm{Y}_j|$ and let $n := \sum_{ij} n_{ij}$. Then, $ARI$ is defined as
\begin{equation}
ARI \!:=\! \frac{ \sum_{ij} \binom{n_{ij}}{2} - \big[ \sum_{i} \binom{|\bm{X}_i|}{2} \sum_{j} \binom{|\bm{Y}_j|}{2} \big] / \binom{n}{2} } { \frac{1}{2} \big[\! \sum_{i} \binom{|\bm{X}_i|}{2} \!+\! \sum_{j} \binom{|\bm{Y}_j|}{2} \! \big] \!-\! \big[\! \sum_{i} \binom{|\bm{X}_i|}{2}\! \sum_{j} \binom{|\bm{Y}_j|}{2} \! \big] / \binom{n}{2} }. \nonumber
\end{equation}
where $|\bm{X}_i|$ is the number of elements in predicted cluster $i$. Similarly for $|\bm{Y}_i|$.

Mutual information~\cite{thomas2006elements} measures the similarity of the two distributions formed by the predicted clustering results $\bm{X}$ and the ground truth clusters $\bm{Y}$ using Kullback-Leibler divergence, which is defined as 
\begin{equation}
MI(\bm{X}, \bm{Y}) := \sum_{ij}\frac{n_{ij}}{n} \log \frac{n\cdot n_{ij}}{|\bm{X}_i||\bm{Y}_j|}. \nonumber
\end{equation} 
The higher the $MI$ value is, the better the clustering results. Since $MI$ is not bounded, in this work, we use the following normalized version of $MI$, i.e., $NMI$:
\begin{equation}
NMI(\bm{X}, \bm{Y}) := \frac{2\cdot MI(\bm{X}, \bm{Y})}{H(\bm{X}) + H(\bm{Y})}, \nonumber
\end{equation}
where $H(\bm{X})$ is the entropy of $\bm{X}$ and defined as 
\begin{equation}
H(\bm{X}) = -\sum_{i} g(\bm{X}_i) \log g(\bm{X}_i), \text{~with~} g(\bm{X}_i) = \frac{|\bm{X}_i|}{\sum_{j}|\bm{X}_j|}. \nonumber
\end{equation}
Similarly for $H(\bm{Y})$. Note that $NMI$ is in $[0, 1]$.

In addition, for the indexing performance, we use an edit distance~\cite{jaro1989advances} to measure how similar two given sequences are by considering the number of transpositions needed to make them identical to each other. Consider a five-cluster case as an example. Suppose that the ground-truth indexing of five clusters is given by $[F1, F2, F3, F4, F5]$. Then, its corresponding ground-truth sequence is $S_Y \!=\! (1,2,3,4,5)$. Also, assuming that the predicted indexing is $[F1, F4, F3, F2, F5]$, we have the predicted sequence as $S_X \!=\! (1,4,3,2,5)$. Thus, in this example, we need one transposition, i.e., to swap $4$ and $2$, to make $S_X$ identical to $S_Y$. Specifically, in this work, we use the following Jaro-Winkler edit distance~\cite{jaro1989advances}: 
\begin{equation}
\text{Edit Distance} := \begin{cases}
0,  &\text{if~} m=0, \\
\frac{1}{3} \Big(\frac{m}{|S_X|} + \frac{m}{|S_Y|} + \frac{m-t}{m}\Big), &\text{otherwise}, \nonumber
\end{cases}
\end{equation}
where $m$ is the number of matching numbers, $t$ is the number of transpositions, and $|S_X|$ and $|S_Y|$ are the lengths of sequences $S_X$ and $S_Y$, respectively.

For all metrics, higher values indicate better performance. 

\begin{figure*}[t]
    \centering
    	\subfloat[Microsoft]{%
        \includegraphics[width=0.23\linewidth]{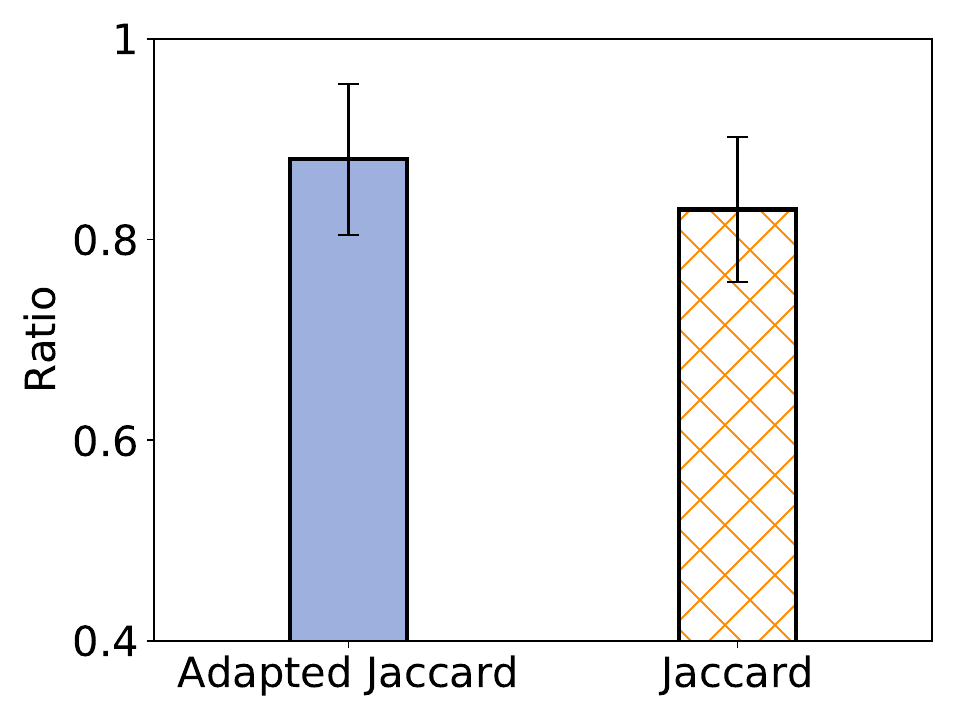}
        }
    \hspace{0.05in}
    \subfloat[Ours]{%
        \includegraphics[width=0.23\linewidth]{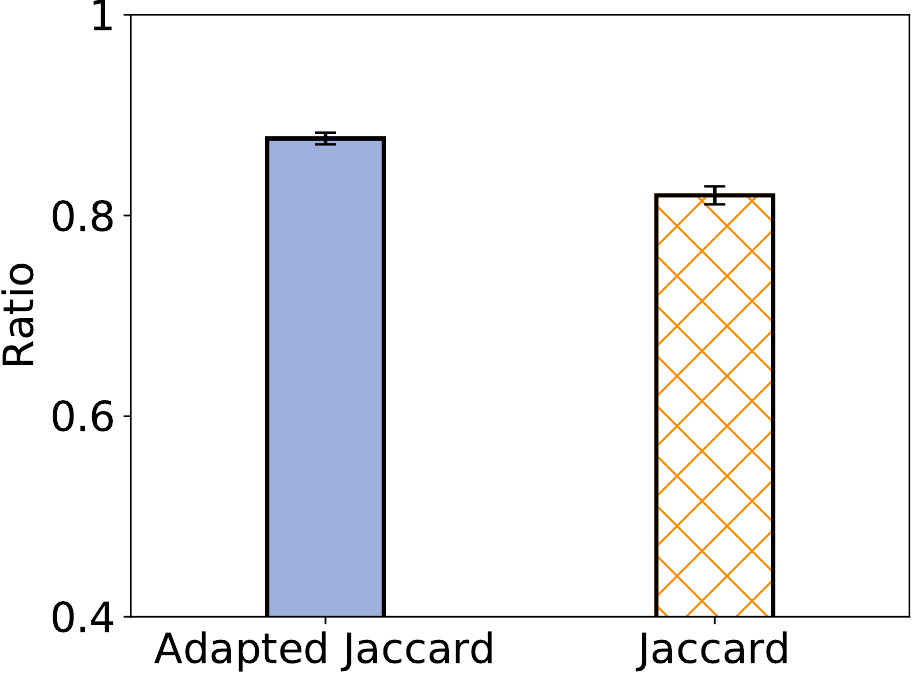}
        }
    \hspace{0.05in}
    \subfloat[Microsoft]{%
        \includegraphics[width=0.23\linewidth]{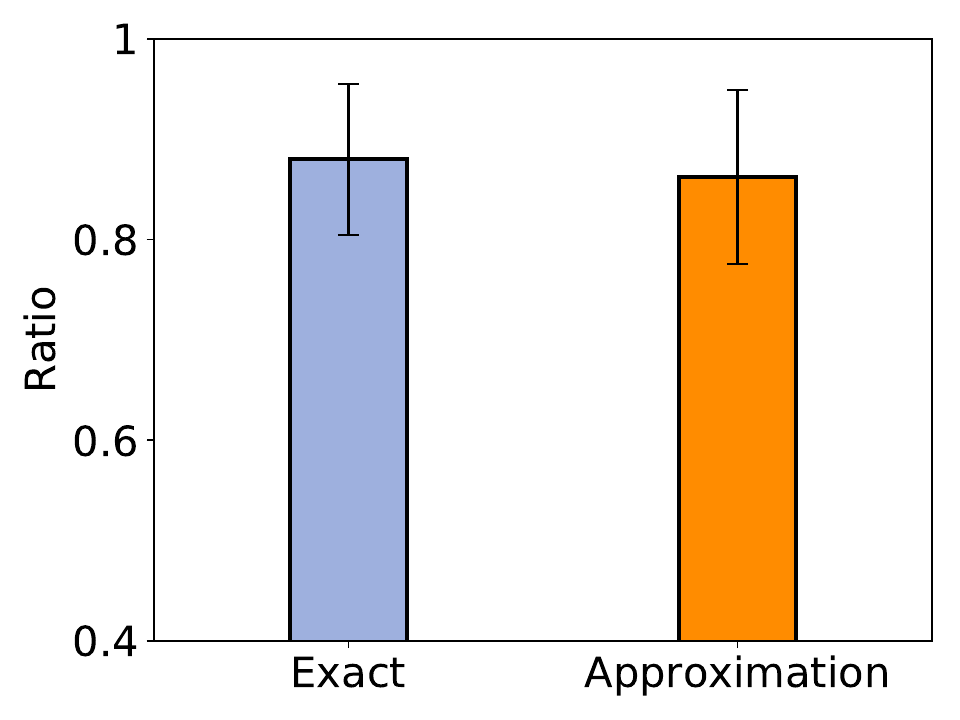}
        }    
    \hspace{0.05in}
    \subfloat[Ours]{%
        \includegraphics[width=0.23\linewidth]{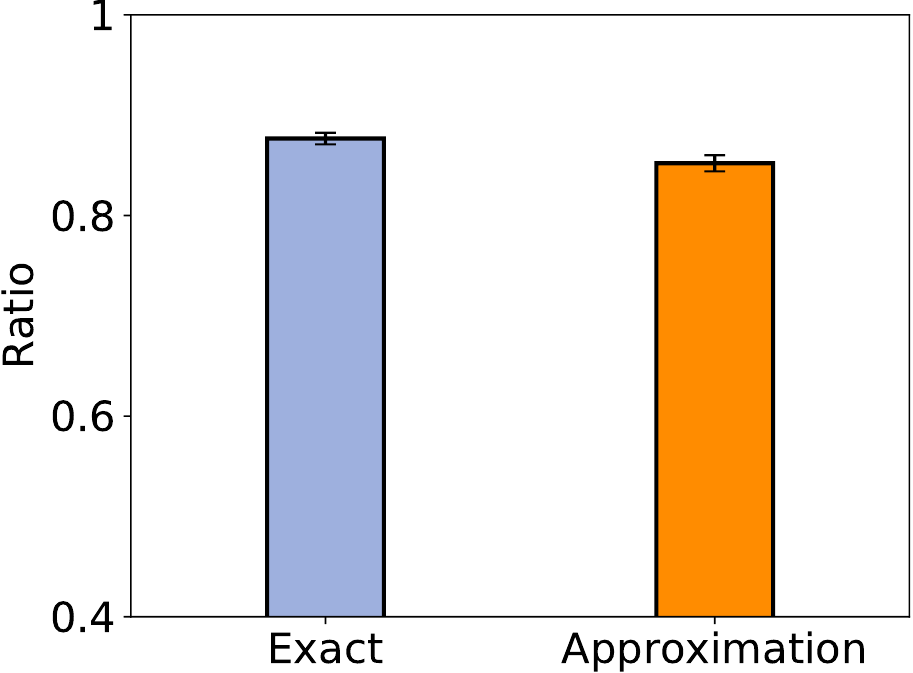}
        }
    	\caption{Ablation study of \n{}. (a) and (b) \n{} (Jaccard similarity coefficient); (c) and (d) \n{} (approximation algorithm for TSP).}
    \label{fig:ablation_cd}
    \vspace{-0.03in}
\end{figure*} 

\begin{figure*}[t]
    \centering
        \includegraphics[width=0.54\linewidth]{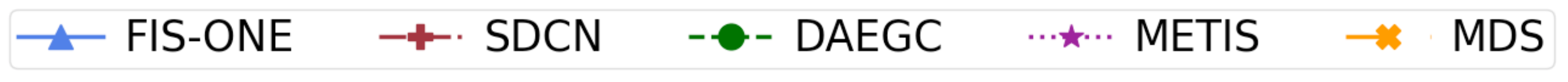}
        \subfloat[Microsoft]{%
        \includegraphics[width=0.24\linewidth]{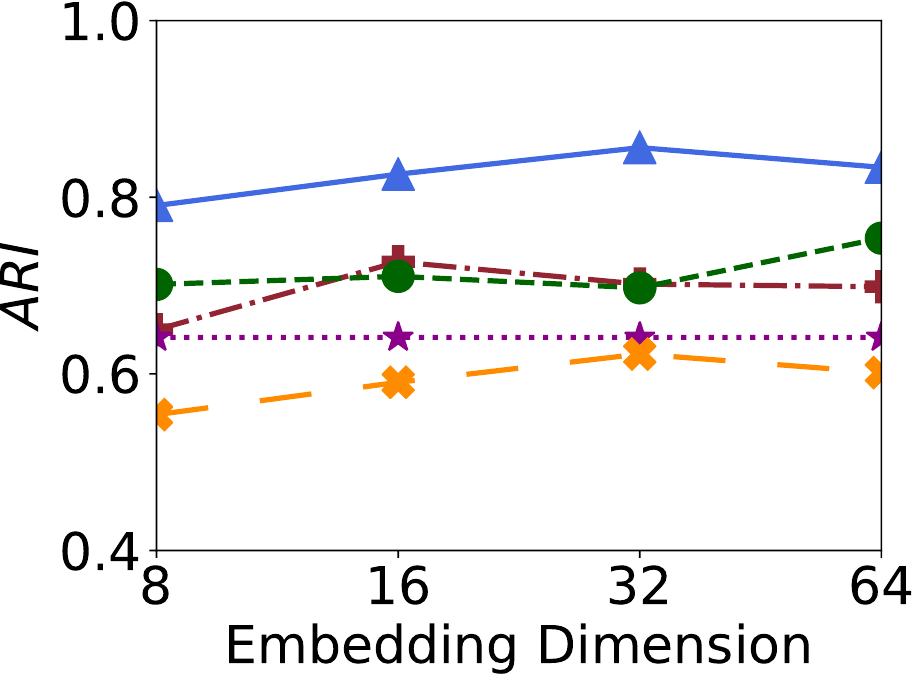}
        }
        \subfloat[Ours]{%
        \includegraphics[width=0.24\linewidth]{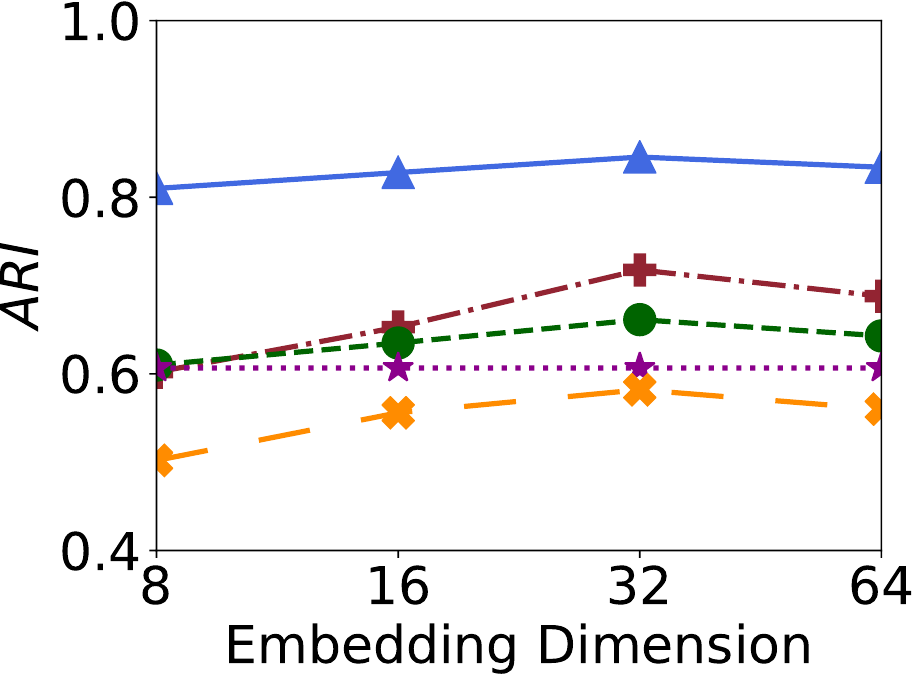}
        }
        \subfloat[Microsoft]{%
        \includegraphics[width=0.24\linewidth]{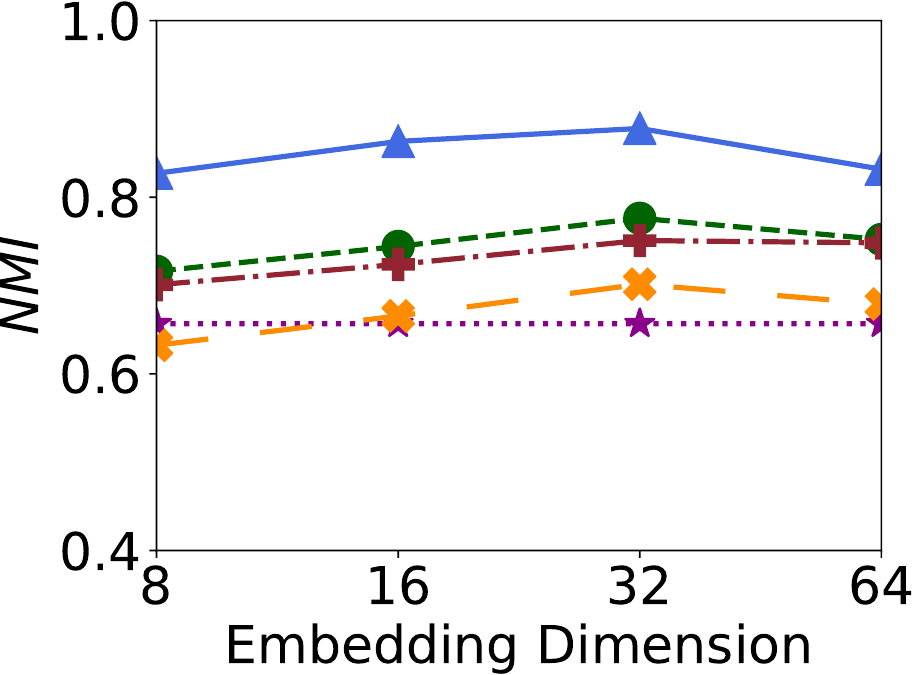}
        }
        \subfloat[Ours]{%
        \includegraphics[width=0.24\linewidth]{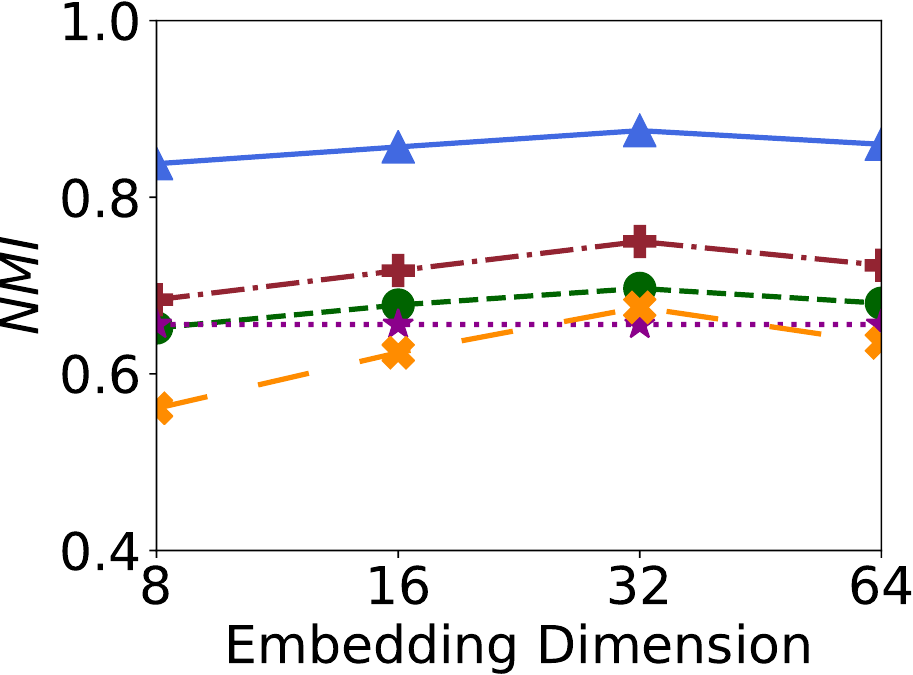}
        }
       \caption{Impact of embedding dimension on the clustering performance. (a) and (b) $ARI$; (c) and (d) $NMI$.}
    \label{fig:ari_nmi_dim}
    \vspace{-0.2in}
\end{figure*} 

\begin{figure}[t]
    \centering
       \subfloat[Microsoft]{%
        \includegraphics[width=0.48\linewidth]{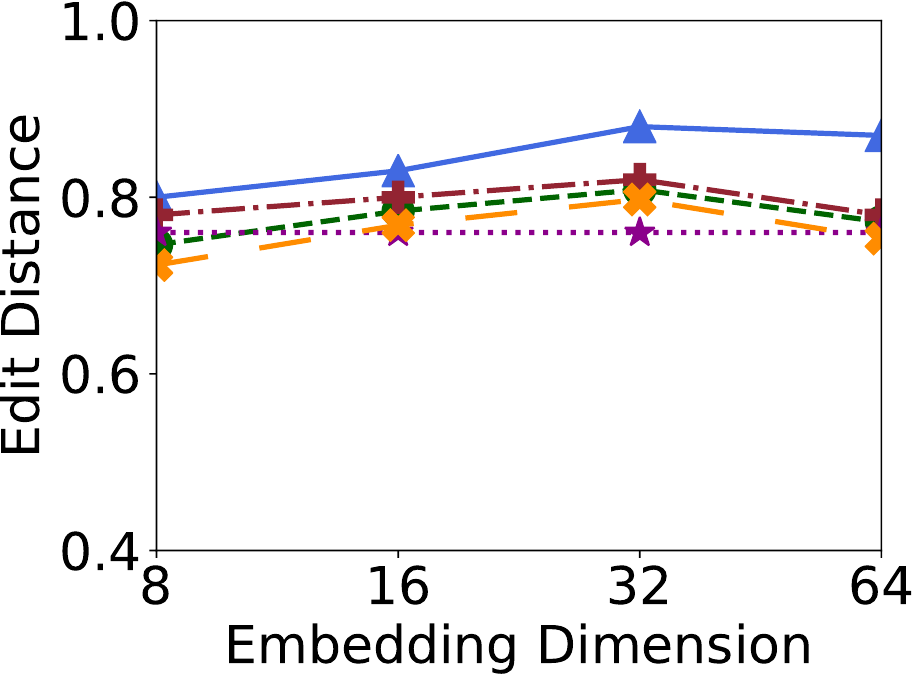}
        }
        \subfloat[Ours]{%
        \includegraphics[width=0.48\linewidth]{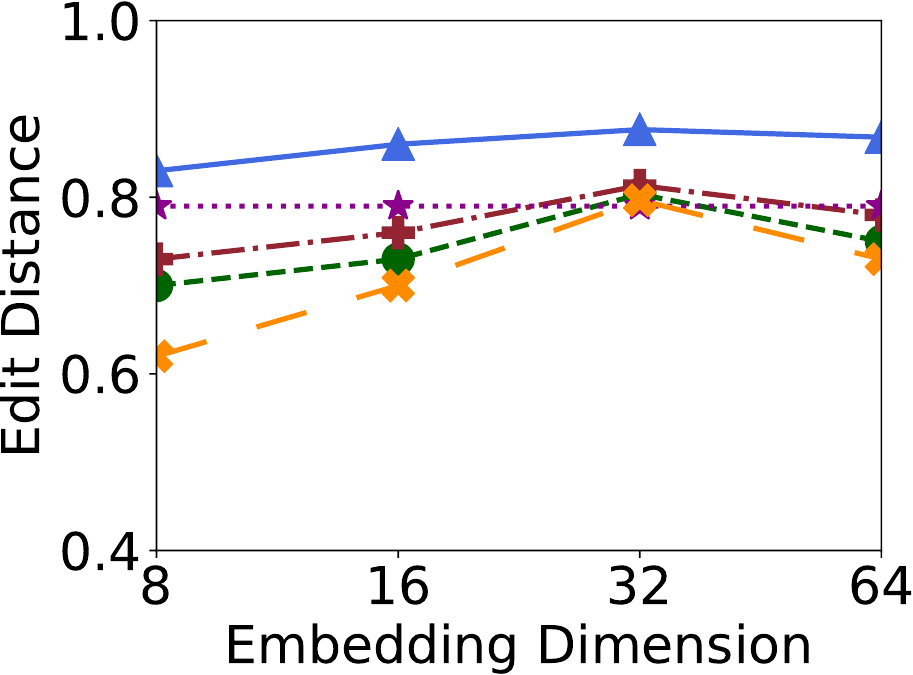}
        }            
       \caption{Impact of embedding dimension on the indexing performance.}
    \label{fig:edit_distance_dim}
    \vspace{0.1in}
\end{figure} 

\subsection{Overall System Performance Comparison}
\label{subsec:eval_overall}

We report, in Table~\ref{tab:results_all}, the clustering and indexing results of \n{} and other baseline algorithms. For clustering, \n{} outperforms SDCN and DAEGC in $ARI$ by more than 20\% and 23\%, respectively. The gain in $NMI$ is also up to 17\% and 25\%, respectively. These all indicate the effectiveness of \nn{} -- our carefully designed representation learning algorithm with an attention mechanism. SDCN obtains clusters in a self-supervised manner by leveraging the centers of clusters. However, the centers estimated during training may not provide good guidance as RF signals on the same floor can even exhibit quite different characteristics, which leads to a multi-modal distribution. DAEGC also suffers from the same problem as the cluster loss that it uses also involves the computation of cluster centroids. METIS does not perform well as the boundary between different clusters of RF signals may not be obvious due to the signal spillover effect. MDS learns the signal embeddings using the matrix of the superset of APs (MACs) to represent RF signals, so it suffers from the missing-value problem (see Figure~\ref{fig:missing_value}).

As shown in Table~\ref{tab:results_all}, \n{} also achieves the best performance in edit distance among all the schemes. This demonstrates that our clusters are well-formed based on \nn{}, and the signal indexing is correctly done based on the optimal solution to the cluster indexing problem, which is transformed into a TSP, where our adapted Jaccard coefficient effectively measures the similarity between clusters. However, the other algorithms show inferior performance in edit distance, which inherits from their low-quality clustering performance.

\subsection{Ablation Study}
\label{subsec:sys_component}

To see the gain that \n{} obtains from the attention mechanism in \nn{}, in Figure~\ref{fig:ablation_ab}(a) and Figure~\ref{fig:ablation_ab}(b), we show the performance of \n{} when \nn{} is used with and without the attention mechanism. As shown in Figure~\ref{fig:ablation_ab}(a) and Figure~\ref{fig:ablation_ab}(b), incorporating edge weights as an attention mechanism in the learning process boosts up the system performance significantly, with up to 80\% improvement in $ARI$, 49\% improvement in $NMI$, and 34\% improvement in edit distance. This is because the edge weight-based attention mechanism correctly incorporates proximity information between different signal samples in learning their vector representations. If two signal samples are collected closely in the physical space, their representations in the latent vector space are also close to each other. Hence, the representations learned with the attention mechanism can be more easily separated across different clusters, leading to better clustering performance.

To study the effectiveness of the hierarchical clustering, we next present the comparison between \n{} with the hierarchical clustering and \n{} with the clustering algorithm being replaced by $K$-means in Figure~\ref{fig:ablation_ab}(c) and Figure~\ref{fig:ablation_ab}(d). We see that the hierarchical clustering performs better than $K$-means when integrated into \n{} (with 4\% improvement in $ARI$, 4\% improvement in $NMI$, and 6\% improvement in edit distance). This is because the hierarchical clustering better handles the signal representations around the boundary as it gradually merges similar representations together from the very beginning. In contrast, $K$-means may not be efficient in differentiating the boundary cases.

We further evaluate the improvement of our adapted Jaccard similarity coefficient over the original Jaccard similarity coefficient and present the results in Figure~\ref{fig:ablation_cd}(a) and Figure~\ref{fig:ablation_cd}(b). Our adapted coefficient achieves higher edit distance with lower standard deviation compared to the original one, meaning that this adapted coefficient better captures the signal spillover effect between floors by considering the appearance frequency of APs (MACs) and thus provides a better similarity measure between signal clusters. As such, the cluster indexing can be done more accurately.

In solving the TSP, the exact algorithm can also be replaced by an approximation algorithm to improve the computational efficiency, possibly at the cost of accuracy loss. We show the results in Figure~\ref{fig:ablation_cd}(c) and Figure~\ref{fig:ablation_cd}(d), where the 2-opt approximation algorithm~\cite{johnson1997traveling} is used to obtain a near-optimal solution to the TSP. We can see that the performance degradation is insignificant ($\sim$ 3\%) by adopting the approximation algorithm. Hence, for tall buildings with many floors, we expect that one can resort to the approximation algorithm as a cost-efficient alternative to achieve accurate floor identification without much performance degradation.

\begin{figure*}[t]
    \centering
    	\subfloat[]{%
        \includegraphics[width=0.24\linewidth]{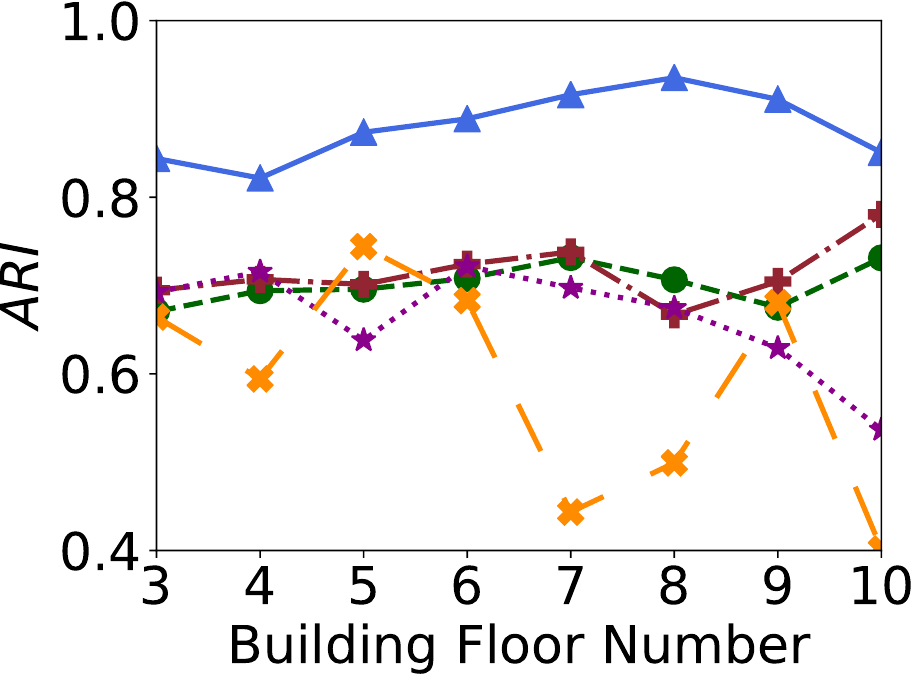}
        }
    \hspace{0.1in}    
    \subfloat[]{%
        \includegraphics[width=0.24\linewidth]{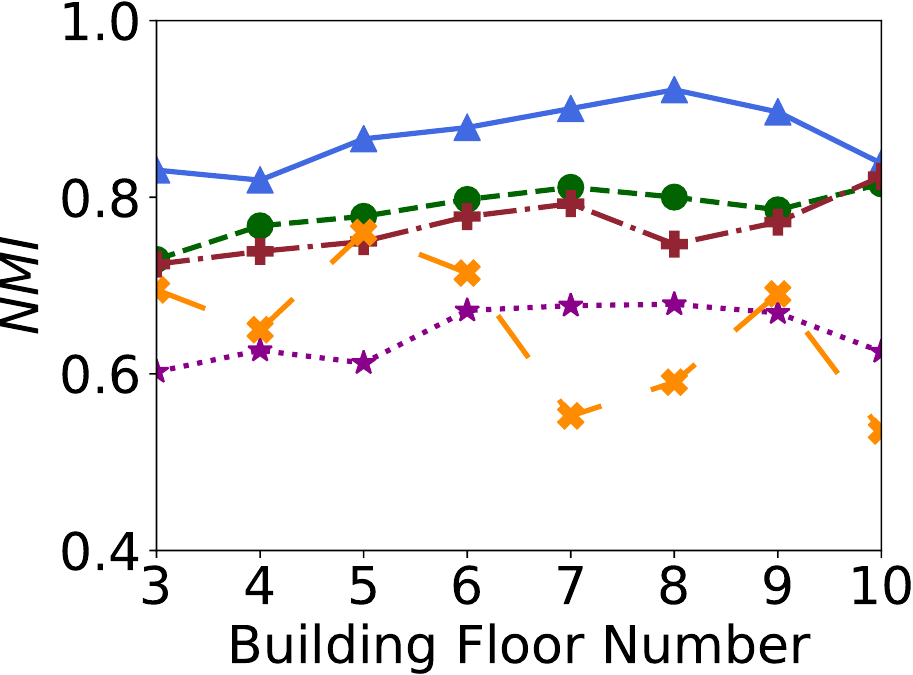}
        }
    \hspace{0.1in}
    \subfloat[]{%
        \includegraphics[width=0.24\linewidth]{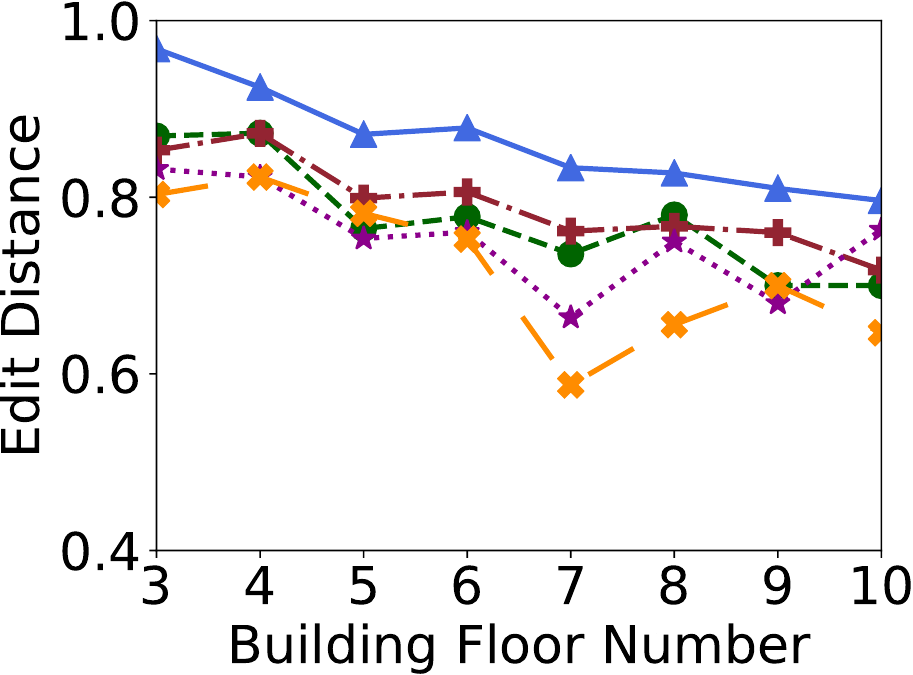}
        }        
    \hspace{0.1in}
    \subfloat{%
        \includegraphics[width=0.1\linewidth]{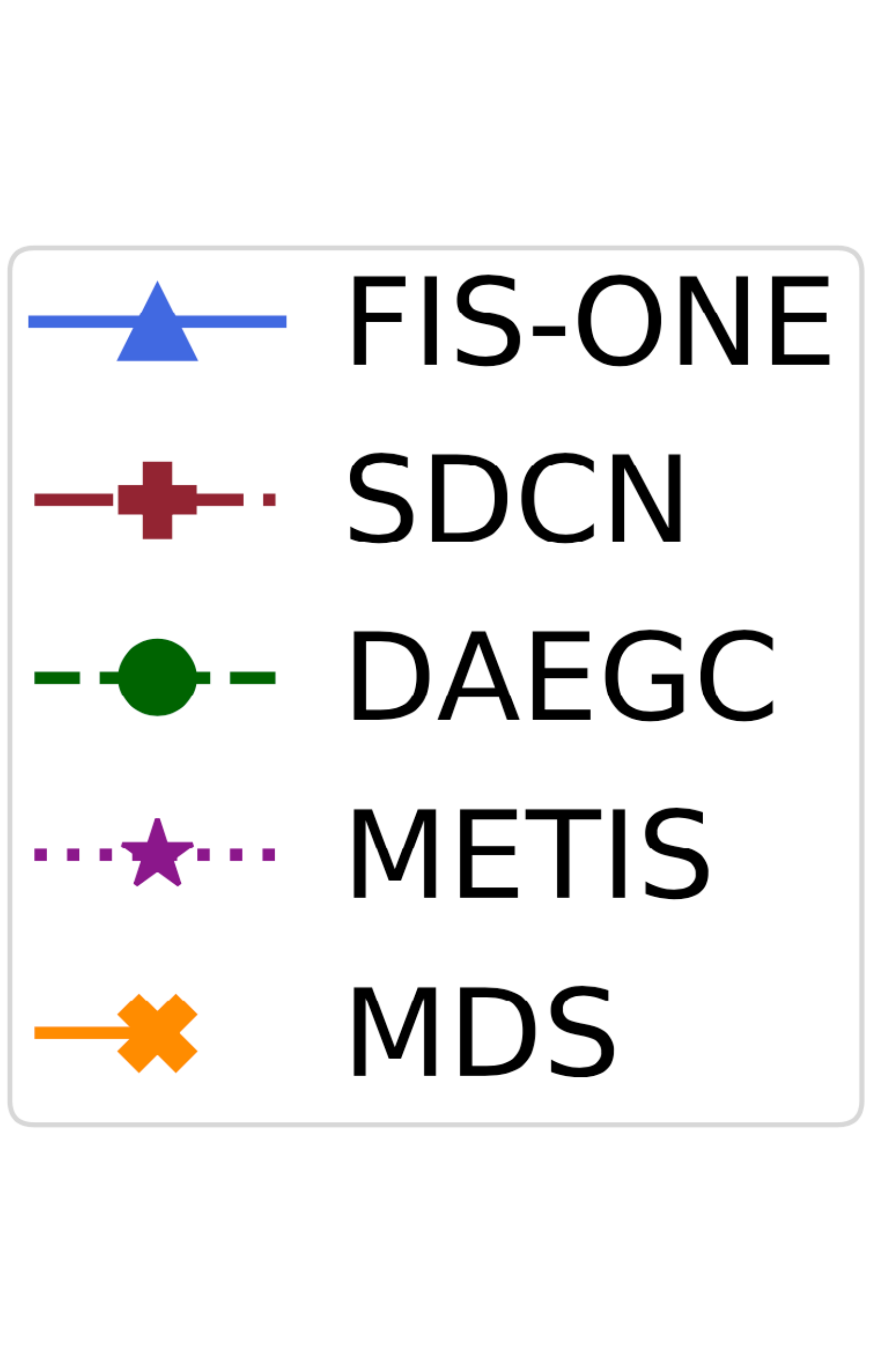}
        }    
    	\caption{Performance of \n{} in different building types (two datasets combined).}
    	\label{fig:param_floors}
    \vspace{-0.15in}
\end{figure*} 

\subsection{System Parameter Study}
\label{subsec:exp_micro}

For practical deployment, we have a wide range of choices for the embedding dimension used in our proposed \nn{} and other baseline algorithms. To check their system sensitivity to this parameter, we vary the embedding dimension from 8 to 64 for each scheme and run the experiments on the two datasets. As presented in Figure~\ref{fig:ari_nmi_dim} and Figure~\ref{fig:edit_distance_dim}, \n{} consistently performs well and better than the other baseline ones across different choices of embedding dimension, meaning that it is robust to changes in the embedding dimension. Note that METIS has no parameter of embedding dimension. We, however, plot its performance for consistency.

We are also interested in evaluating how \n{} performs for different building types, i.e., buildings of different floor numbers. Hence, we summarize the statistics in Figure~\ref{fig:param_floors}. We see that \n{} performs well for all building types with small fluctuations, and it is consistently better than the other baseline algorithms. The performance of \n{} and other baseline algorithms overall fluctuates a bit more for taller buildings. This is because there is a fewer number of such buildings (see Figure~\ref{fig:building_num}), exhibiting larger variations due to a smaller sample size. Nonetheless, \n{} still performs well under such cases, which again verifies its effectiveness.

\section{Discussion} 
\label{sec:discuss}
We discuss here the feasibility of using only one labeled signal sample from an \emph{arbitrary} floor instead of the bottom (or top) floor. So far we have assumed that the labeled sample is collected on the bottom (or top) floor, which is used as an indicator of the starting point for the TSP. It ensures that only one path with the maximum sum of adapted Jaccard similarity coefficients along the path can be obtained. Thus, we can index the clusters correspondingly, as explained in Section~\ref{sec:cluster_index}.

\begin{figure}[t]
    \centering
    \includegraphics[width=0.9\linewidth]{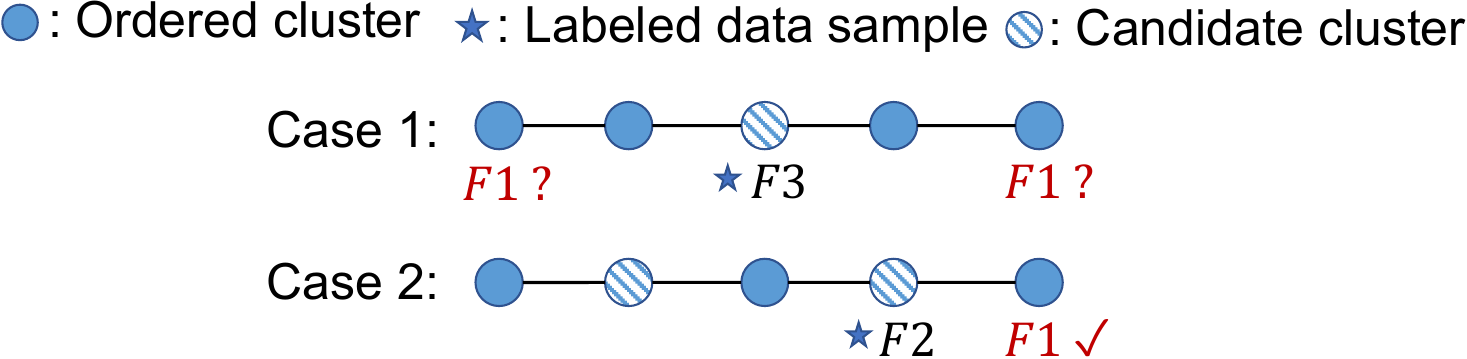}
    	\caption{\textbf{Case 1}: the labeled sample is from the middle floor in which case we cannot do the prediction. \textbf{Case 2}: the labeled sample is from other floors in which case we predict which candidate cluster is closer to the labeled sample.}
    	\label{fig:discuss_random_label}
    \vspace{0.05in}
\end{figure} 

Now, we explain how we can relax the assumption such that the labeled sample can be collected from an \emph{arbitrary} floor. We first do not consider the labeled signal sample in the clustering process after its vector representation is obtained. Since there is no fixed starting point for the TSP, we solve the TSP with all possible starting points, e.g., leading to $N$ orderings for a building of $N$ floors. From these orderings, we pick out the one with the maximum sum of adapted Jaccard similarity coefficients and use the ordering for cluster indexing. However, there are two cases to consider. 

\textbf{Case 1}: The building has an odd number of floors, and the labeled sample is collected from the middle floor. For instance, as shown in Figure~\ref{fig:discuss_random_label}, there are five floors in the building, and the labeled sample is collected from the third floor. Hence, it is not possible to index the ordering, as there is no indicator which side of the sequence contains the starting floor. 

\textbf{Case 2}: For all the other scenarios, given a labeled sample, we can always find two candidate clusters to locate the labeled signal sample, as shown in Figure~\ref{fig:discuss_random_label}. Then, we ``predict" which candidate cluster the labeled sample belongs to by finding the cluster that is \emph{closer} to the labeled sample. The distance between cluster $i$ and the vector representation of the labeled sample, say, $\bm{r}$, is calculated as 
\begin{equation}
d(\bm{r}, \bm{C}_i) := \sum_{\bm{r}' \in \bm{C}_i} \frac{\| \bm{r}'-\bm{r} \|_2}{|\bm{C}_i|}. \nonumber
\end{equation}
In other words, we calculate the averaged pairwise distance between the vector representations in $\bm{C}_i$ and the representation of the labeled sample. 

\begin{figure}[h]
\vspace{-3mm}
    \centering
    \subfloat[]{%
        \includegraphics[width=0.48\linewidth]{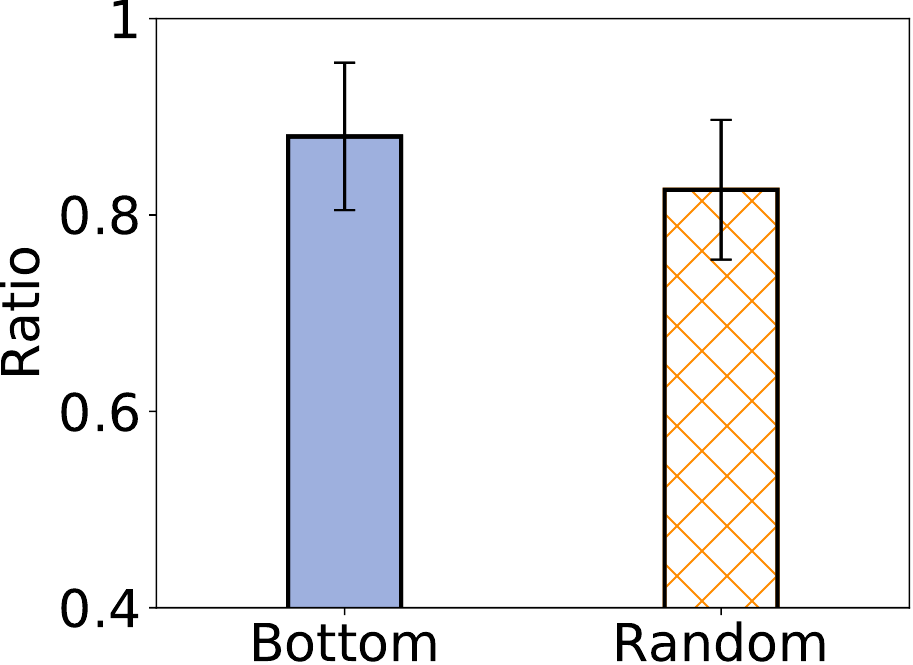}
        }
    \subfloat[]{%
        \includegraphics[width=0.48\linewidth]{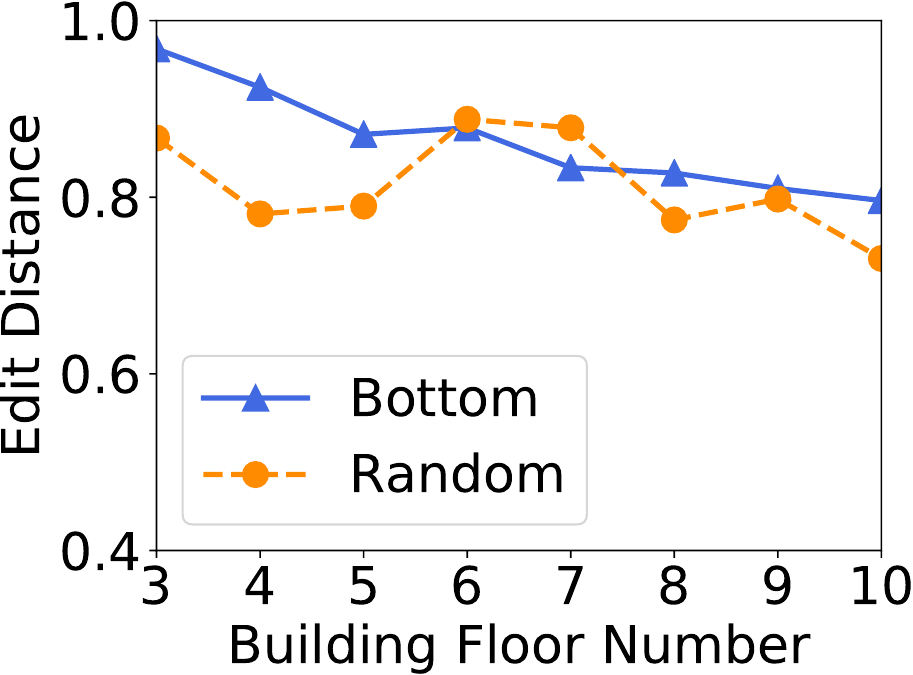}
        }
    	\caption{Performance comparison when using a labeled sample from bottom floor (\n{}) and a labeled sample from a random floor (two datasets combined). (a) Overall edit distance comparison. (b) Building-wise edit distance comparison.}
    	\label{fig:label_random}
     \vspace{0.00in}
\end{figure}

To check the feasibility of this approach, we conduct an experiment on the two datasets with a labeled sample obtained from a random floor in Case 2. The experiment is repeated for ten times, and the average results are presented in Figure~\ref{fig:label_random}. We see that \n{} still performs well without much performance degradation ($\sim $7\%). 

\section{Conclusion}
\label{sec:conclude}
We proposed \n{}, a floor identification system for crowdsourced RF signals with only one labeled signal sample. Based on the observation of signal spillover, \n{} clusters the RF signals effectively and indexes the clusters accurately. As an integral component of \n{}, \nn{} enables efficient representation learning for a large number of RF signals on a (possibly dynamic) graph. Extensive experiment results on Microsoft's open dataset and in three large shopping malls validated the effectiveness of \n{} and demonstrated its superior performance over baseline algorithms (with up to 23\% improvement in $ARI$ and 25\% improvement in $NMI$). We also discussed how \n{} can be extended to the case when the one labeled signal sample comes from an \emph{arbitrary} floor. We believe that we have taken a first step towards \emph{unsupervised} floor identification for crowdsourced RF signals.

\bibliography{ref_short}
\bibliographystyle{IEEEtran}

\end{document}